\pdfoutput=1

\documentclass[3p,times]{elsarticle}

\usepackage{ecrc}

\usepackage{amsmath,amsfonts,amssymb,esint}
\usepackage{algorithm}
\usepackage{algorithmic}
\usepackage{graphicx}
\usepackage{enumitem}
\usepackage[lofdepth,lotdepth]{subfig}

\usepackage[utf8]{inputenc}
\usepackage[T1]{fontenc}
\usepackage[USenglish]{babel}

\usepackage{todonotes}
\presetkeys{todonotes }{author=JR, inline, color=green!50!white, bordercolor=blue }{}

\volume{00}

\firstpage{1}

\journalname{Journal of Computational Science}

\runauth{Couteyen Jean Marie}

\jid{procs}

\jnltitlelogo{Procedia Computer Science}

\CopyrightLine{2017.}{Licensed under the CC-BY-NC-ND 4.0 license. (http://creativecommons.org/licenses/by-nc-nd/4.0/) \\ DOI: 10.1016/j.jocs.2017.03.008 }

\usepackage{amssymb}

\usepackage[figuresright]{rotating}

\begin{document}

\begin{frontmatter}

\dochead{}

\title{Design and Analysis of a Task-based Parallelization over a Runtime System of an Explicit Finite-Volume CFD Code with Adaptive Time Stepping}

\author[agi,asl]{Jean Marie Couteyen Carpaye}
\author[inria]{Jean Roman}
\author[asl]{Pierre Brenner}

\address[agi]{Airbus Group Innovations, 12 Rue Pasteur, 92150 Suresnes, France}
\address[asl]{Airbus Safran Launchers, 51-61 Route de Verneuil, 78130 Les Mureaux, France}
\address[inria]{Inria, Bordeaux INP, CNRS (LaBRI UMR 5800), 33400 Talence, France}

\address{}

\begin{abstract}
FLUSEPA\footnote{Registered trademark in France No. 134009261} is an advanced simulation tool which performs a large panel of aerodynamic studies. It is the unstructured finite-volume solver developed by Airbus Safran Launchers company to calculate compressible, multidimensional, unsteady, viscous and reactive flows around bodies in relative motion. The time integration in FLUSEPA is done using an explicit temporal adaptive method. The current production version of the code is based on MPI and OpenMP.
This implementation leads to important synchronizations that must be reduced.
To tackle this problem, we present the study of a task-based parallelization of the aerodynamic solver of FLUSEPA using the runtime system StarPU and combining up to three levels of parallelism. We validate our solution by the simulation (using a finite-volume mesh with 80 million cells) of a take-off blast wave propagation for Ariane 5 launcher.
\end{abstract}


\end{frontmatter}

\section{Introduction}

For industrial applications of numerical simulation, the most common
architecture is nowadays clusters composed of SMP nodes of multicore
processors. To develop parallel codes on those machines, a common way
is to rely on MPI \cite{forum_mpi:_1994}. While it is possible to use
only one core per process and to rely only on Flat-MPI, this approach
does not generally scale and it is even worse for codes with a large
potential imbalance during execution. A common way to reduce the
number of processes while using the same number of cores consists in
using OpenMP \cite{dagum_openmp:_1998} inside the SMP nodes, leading
to a two level parallelism.

Another problem is the increasing of the heterogeneity of
architectures. Accelerators (e.g. GPGPU, Xeon Phi) are now available
and the design of efficient industrial applications exploiting
distributed heterogeneous systems with non uniform memory accesses is
a complex challenge at large scale.
Therefore, applications tend to evolve slowly compared to
architectures and achieving performance with a new architecture is a
time-consuming task for the developers.

Task-based programming is a good candidate to deal with those issues:
describing the problem in a generic manner as a DAG of tasks allows
more potential flexibility to exploit the architectures. A runtime
system is then in charge of mapping tasks among computational
resources (CPU cores and/or accelerators) and of managing memory
transfers.
By using a powerful abstraction of parallel codes and an efficient
runtime system, one can expect to achieve performance quickly on
different kinds of architectures (performance portability issue
\cite{AugThiNamWac11CCPE},\cite{bosilca_parsec:_2013}).
Moreover, this approach allows to share efficiently resources among different 
parallel libraries or codes~\cite{hugo:hal-00824514} while enabling a fine
description of the interactions between them. This is a key advantage when tackling
complex multiphysics and/or multiscale simulations that imply code coupling.

The aerospace industry faces a lot of problems involving complex
unsteady computations for which actual experiences are not doable:
e.g. take-off blast wave propagation, stage separation, stability at
reentry into Earth atmosphere. Those problems are time-dependent and
involve strong shocks.

In this paper, we describe the ``taskification" using the runtime
system StarPU \cite{AugThiNamWac11CCPE} of the aerodynamic solver of
the FLUSEPA code \cite{brenner1991three} which is an MPMD MPI/OpenMP
code.
This aerodynamic solver uses a finite-volume (FV) discretization and
an explicit temporal adaptive time stepping scheme.
This kind of scheme is well suited for our class of problem because it
is conservative and consistent in time
(\cite{kleb_temporal_1992,lohner_finite_1985,pervaiz_temporal_1988}). The
method is designed to minimize the computational cost, but it
introduces several difficulties for an efficient parallelization
(synchronizations, load balancing problems). In this study, we show
how a task-based solution over a runtime system can be efficient to
tackle these problems.

A preliminary version of this work appeared in \cite{mypdsec}.
In this paper, we give first a more detailed analysis of our new parallel version
of the aerodynamic solver in shared memory. In particular, the experimental section
now contains an analysis at two different iterations of the global computation and with
different cell distributions according to the temporal levels.
A second contribution consists in the validation of our approach in a distributed parallel
context. We performed a large simulation using a finite-volume mesh with 80 million cells
and by using 28 nodes (560 cores) on a parallel cluster.

The paper is organized as follows. Section \ref{mmrv} briefly presents
the main computational methods used in FLUSEPA and the existing
MPI/OpenMP code. Section \ref{starpu} presents the runtime system
StarPU and Section \ref{impl} focuses on the task parallelization of
the aerodynamic solver. Section \ref{xp} presents an experimental
study from an industrial test case, the take-off blast wave
propagation for Ariane 5 launcher. Finally, Section \ref{conclusion}
gives some conclusions and perspectives for this work.

\section{Main numerical methods used in FLUSEPA}
\label{mmrv}

In this Section, we present the main numerical methods involved in the
general computation performed by the FLUSEPA code with a more detailed
algorithmic description of the aerodynamic solver on which we will
focus in the rest of the paper.

\subsection{Finite-volume aerodynamic solver}
\label{fvas}

The finite-volume method \cite{versteeg_introduction_2007} is the
discretization technique used by FLUSEPA. It is mainly an integral
formulation of the conservation laws which are discretized directly in
the physical space.
In FLUSEPA, we look for the numerical solution of the compressible
Navier-Stokes equations in Reynolds-Averaged form (RANS equation for
3D compressible unsteady and reactive flows involving bodies in
relative motion \cite{germano1999rans}) that can be written as : 

\begin{equation}
\frac{d }{\partial t}\iiint \limits_{\Omega_{CV}}{\mathbf{w}}d\Omega
=-\oiint \limits_{A_{CV}}{\mathbf{F}}\mathbf{n}dS
+\iiint \limits_{\Omega_{CV}}\mathbf{S}d\Omega
\label{integrale}
\end{equation}

where $\Omega_{CV}$ is a fixed control volume (3D cell) with boundary
$A_{CV}$ (2D faces), $\mathbf{n}$ is the outer-oriented unit normal,
$\mathbf{w}$ is the conservative variable vector, $\mathbf{F}$ is the
flux density and $\mathbf{S}$ is the source term vector. 
So, the solver mainly manipulates cells and faces. Field values
(e.g. pressure, temperature) are computed for cells and flows are
evaluated between faces of cells (refer to
\cite{versteeg_introduction_2007} for more details).
The main advantage of this discretization method is its conservativity
leading to a very close approximation to the physics ; moreover, by
using an upwind methodology (Godunov’s method
\cite{godounov_resolution_1979}) for solving the Riemann problem, one
can compute accurately numerical fluxes even when strong shock waves
are involved.

\begin{figure}[!h]
  \centering
  \includegraphics[width=0.24\textwidth]{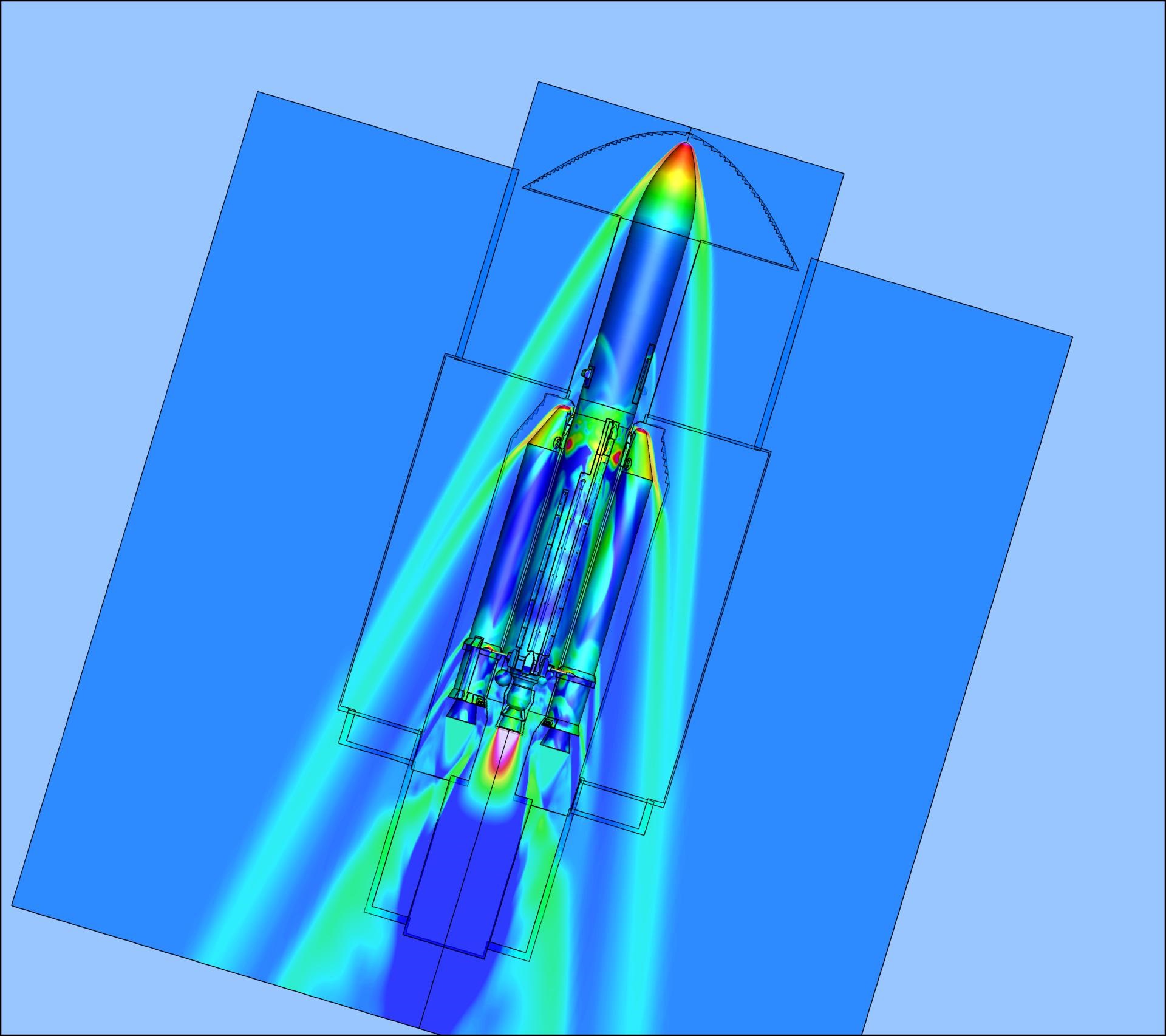}
    \includegraphics[width=0.24\textwidth]{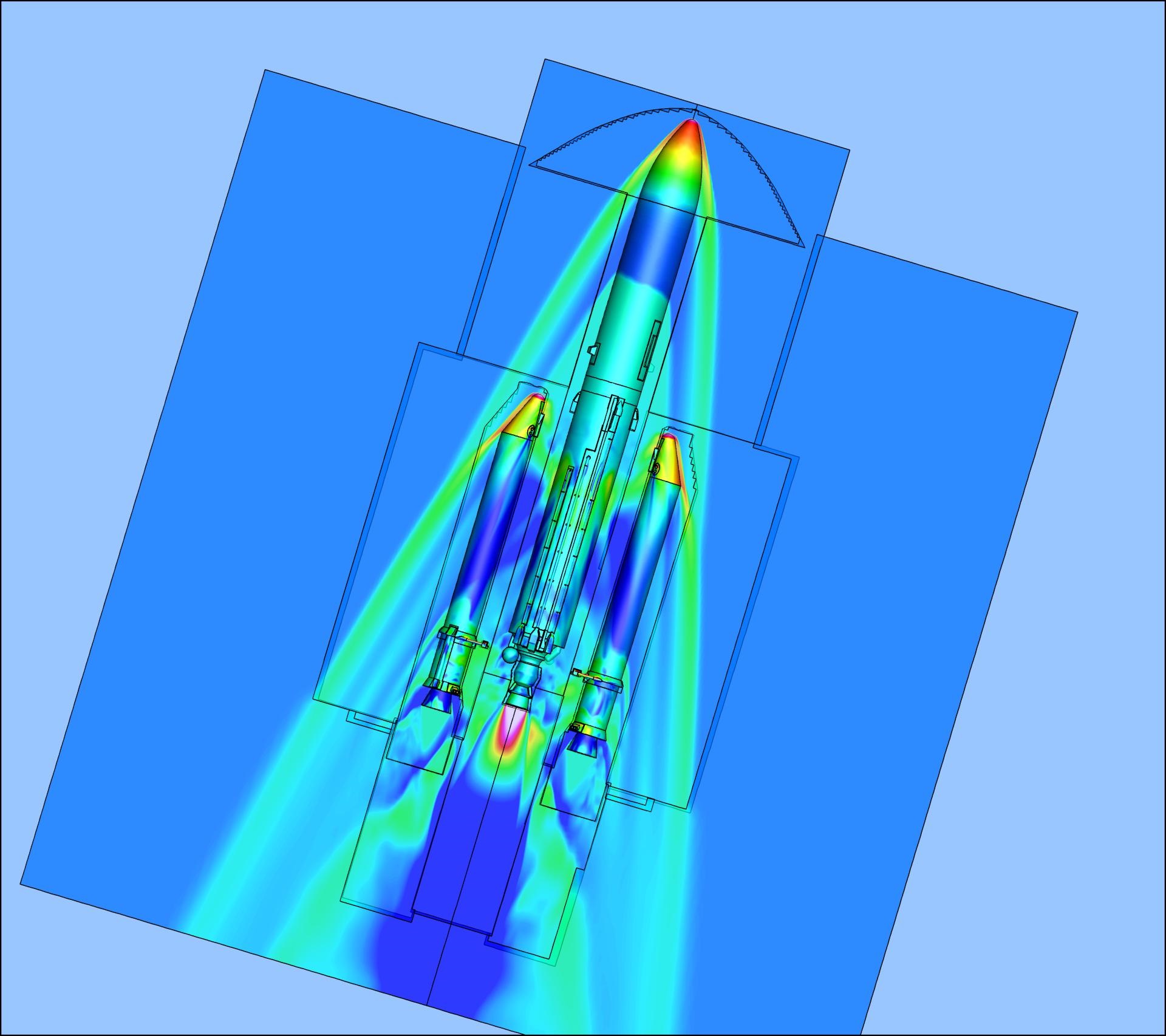}
  \includegraphics[width=0.24\textwidth]{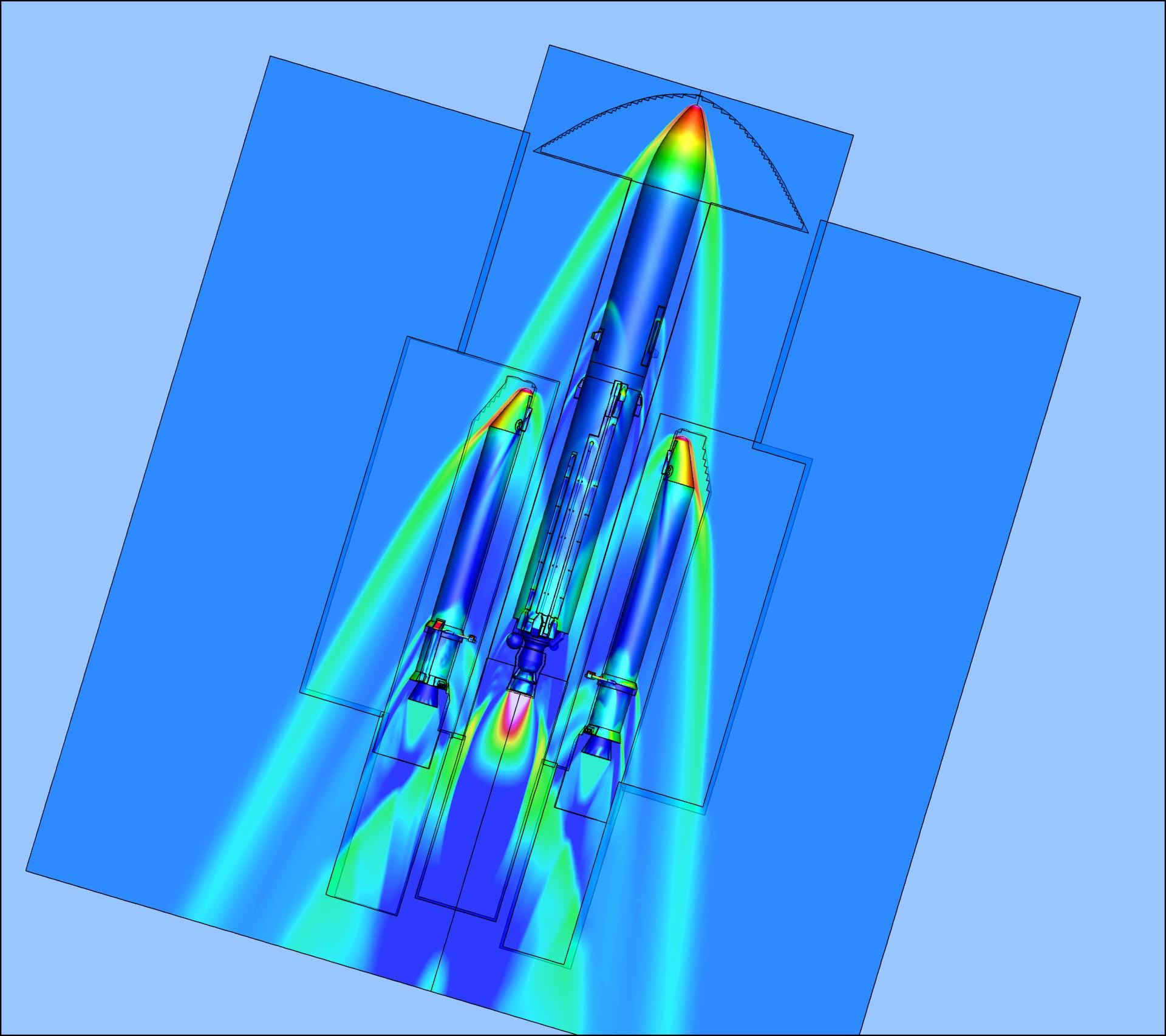}
  \includegraphics[width=0.24\textwidth]{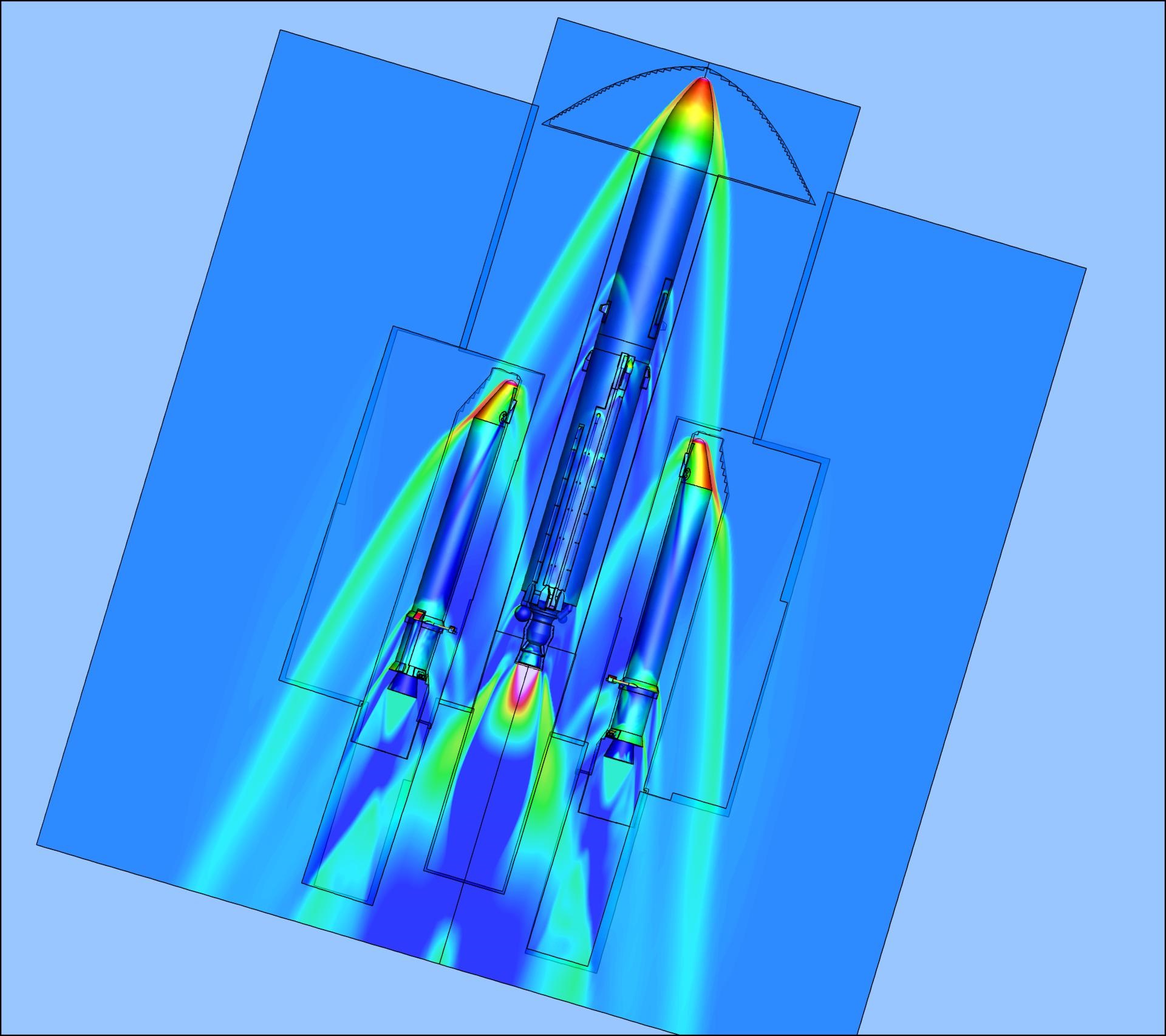}    
  \caption{Booster separation.}  
  \label{a5sep}
\end{figure}

To take into account the mesh motion during stage separation, an ALE
formulation is used \cite{hirt_arbitrary_1974} and when computing
fluxes, the intrinsic velocity of each cell face is taken into
account.
The computation mesh is obtained by a geometric intersection of
several meshes (one for each body in relative motion).
This conservative technique does not make use of numerical
interpolation unlike standard CHIMERA methods
\cite{gillyboeuf1995two,kao1994grid}.
The volume displacement is also integrated in the associated
computational scheme and this is performed by  applying the Geometric
Conservation Law (GCL) \cite{brenner1991three}. This conservation law
is useful in order to 
transform the geometrical intersection problem into an evolution
one. When the relative motion between two meshes is “slow”, computing
the intersection is not mandatory : the evolution of the cell volumes
can be evaluated precisely with the desired level of accuracy through
the use of the GCL.
Figure \ref{a5sep} shows a separation stage computation: each
booster and the main stage are meshed separately. 

The two main computation steps performed in FLUSEPA at each general
iteration, the aerodynamic solver and the mesh intersection
computations, are linked by a kinematic computation. During
aerodynamic computations, external forces apply to the bodies. From
the computation of these loads, a kinematic is obtained. If the bodies
moved sufficiently since the last mesh intersection computation, the
meshes are moved according to the computed kinematic. Otherwise, the
GCL is used in order to take into account the
displacement without computing a full intersection between meshes and
the kinematics is conserved.
This general computational framework is summed up in Algorithm
\ref{general_scheme}.

\begin{algorithm}
\caption{General iteration}
\begin{algorithmic}[1]
\STATE Aerodynamic solver computation
\STATE Computation of a new\_kinematic
\IF{important motion since last mesh intersection}
	\STATE Body displacement (new\_kinematic)
	\STATE Intersection computation
\ELSE
        \STATE Body displacement computation using the GCL
\ENDIF
\end{algorithmic}
\label{general_scheme}
\end{algorithm}

\subsection{Temporal adaptive explicit solver}

The aerodynamic solver used is explicit and based on a temporal adaptive time stepping scheme. When using an explicit temporal formulation, the maximum allowable time step of a cell is given by its CFL number which depends mostly of the volume of the cell. For an explicit solver, the CFL must be inferior to 1. In classical explicit solvers, the time step is determined by the slowest cell (the cell which has the slowest time step while respecting the CFL condition).

Because we consider complex real size problems, the mesh resolution is not uniform, so using the lowest physical time step would be very penalizing for larger cells. The temporal adaptive algorithm allows to compute each cell near its maximum allowable explicit time step, ranking them in levels; these temporal levels $\tau$ are numbered from $0$ to a given value $\theta$.

\begin{algorithm}

\caption{Temporal adaptive time stepping scheme in FLUSEPA: one iteration of the aerodynamic solver}
\begin{algorithmic}[1]
\STATE Time step computation
\STATE Classification of every cell inside a temporal level
\STATE \textit{Temporal adaptive loop:}
\FOR{subiteration=1 \TO $2^\theta$}
	\STATE $\tau=0$
	\FOR{$tmp=1$ \TO $\theta$}
		\IF{$(mod(subiteration-1,2^{tmp})==0)$} \STATE $\tau=tmp$ \ENDIF
	 \ENDFOR
	\IF{subiteration$>$1}
		\STATE Intensive Correction (\{0...$\tau$\})
		\STATE Intensive interpolation ($\tau+1$)
	\ENDIF
	\STATE \textit{Predictor:}
	\STATE Gradient computation (\{0...$\tau$\})
	\STATE Limitation and flow reconstruction (\{0...$\tau$\})
	\STATE Flow repositionning ($\tau+1$)
	\STATE Riemann Solver (\{0...$\tau$\})
	\STATE Flow sum on cells (\{0...$\tau$\})
	\IF{$\tau \not= \theta$}\STATE Intensive repositionning ($\tau+1$) \ENDIF
	\FOR{$\tau'=\tau$ \TO $0$}
		\STATE \textit{End of predictor for $\tau'$:}	
		\STATE Extensive prediction ($\tau'$)
		\STATE Intensive prediction ($\tau'$)
		\STATE \textit{Corrector:}
		\STATE Gradient computation ($\tau'$)
		\STATE Limitation and flow  reconstruction ($\tau'$)
		\STATE Flow interpolation ($\tau'$)
		\STATE Riemann Solver ($\tau'$)
		\STATE Flow sum on cells ($\tau'$)
		\STATE Extensive Correction ($\tau'$) 
		\STATE Intensive interpolation ($\tau'$)
	\ENDFOR
\ENDFOR
\end{algorithmic}
\label{ta_algo}
\end{algorithm}

The temporal adaptive method is described in Algorithm \ref{ta_algo}. At line 1, the maximum allowable time step is computed for each cell. 
The slowest cell is also found and defines $\Delta t$, the minimum time step in the computation. At line 2, according to $\Delta t$ and its maximum allowable time step, each cell is classified inside a temporal level. Inside a temporal level $\tau$, cells are computed at the same time step which is $2^\tau*\Delta t$.

An iteration of the algorithm is composed of multiple subiterations. There are $2^\theta$ subiterations, $\theta$ being the level of the fastest cells which are the ones that need only one subiteration to reach the time of the end of the iteration.

\begin{figure}[!h]
  \centering
  \includegraphics[width=0.99\textwidth]{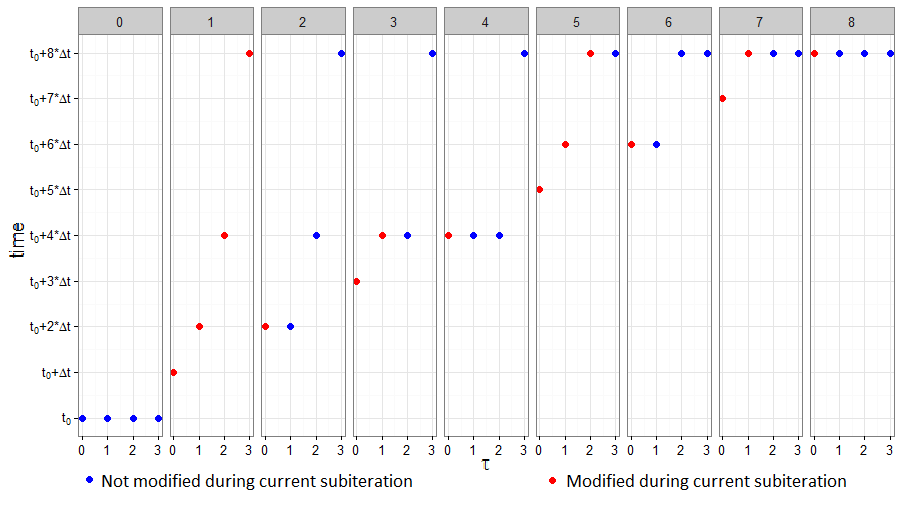}
  \caption{Time reached after each subiteration for each level $\tau$ ($\theta=3$; 8 subiterations).}  
  \label{ta_evo}
\end{figure}

The levels $\tau$ that are computed are determined by lines 5 to 9. 
For example, with $\theta=3$, $\tau$ will take successively the
following values : 3, 0, 1, 0, 2, 0, 1, 0. Figure \ref{ta_evo} shows how
level $\tau$ evolves after each subiteration in this case; the
evolving levels $\tau$ are in red. The physical time reached after one
iteration with $\theta=3$ is equivalent to 8 iterations with a global
time step. 

To stay consistent in time, the computations have to be performed in a specific order. When computing a flow between two cells, they must be at the same time. 
The cells can only have neighbor cells of the same level $\tau$ or with the level values $\tau-1$ or $\tau+1$. If cells are near other cells with a different temporal level, they are positioned at a time that will ensure a consistent computation (lines 18, 22, 30, 34) \cite{kleb_temporal_1992}. 
In this way, the computation order is strict and each temporal level is integrated at a specific moment.

The interest of the method depends strongly of the distribution in temporal levels. Let us denote by $\Omega(\tau)$ the set of cells at temporal level $\tau$ and by $\Omega$ all the cells in the global domain. 

The computational cost of a temporal level $\tau$ can be estimated by
$ C(\tau)=2^{\theta-\tau}*|\Omega(\tau)|$
and the cost of an iteration described by Algorithm \ref{ta_algo} is
$\sum_{\tau=0}^{\theta}C(\tau).$

If we compare the computational cost needed to reach the same time with a global time step ${2^\theta*|\Omega|}$, the cost ratio is
$$\frac{2^\theta*|\Omega|}{\sum_{\tau=0}^{\theta}2^{\theta-\tau}*|\Omega(\tau)| }$$
which is superior to 1.

However, this estimation is an upper bound because of the overhead induced by the temporal adaptive method. Several interpolations are not taken into account in this cost ratio, while their importance increases with the number of temporal levels. More temporal levels imply more cells concerned by interpolation and the cells with a small time step are interpolated more often.

\subsection{MPI-OpenMP version of FLUSEPA}

FLUSEPA uses several kinds of processes to handle the numerical
coupling between aerodynamic computations and body movements. Three
kinds of processes are used: one process is in charge of coordinating
the simulation, some other processes are used to compute the
aerodynamic solution and the last kind of processes is dedicated to
compute mesh intersections \cite{brenner_simulation_1998}. 
This version allows to compute intersections during aerodynamic 
computations since it is possible to rely on the GCL to have accurate
 simulation without computing every intersections. So, an intersection in
 the future of the simulation is computed thanks to an extrapolation of
 the kinematic and then applied when this time is reached by the 
 aerodynamic solver.
In the following of the paper, we focus on the aerodynamic solver
computations of FLUSEPA simulations.\\

The code uses unstructured meshes in order to take into account complex geometries for interstages. The elements manipulated by the solver are the cells and the faces between them.
The current parallel version is based on a two-level parallelism: MPI \cite{forum_mpi:_1994} processes associated with a spatial decomposition and OpenMP \cite{dagum_openmp:_1998} parallelism inside them.

Making a MPI version of a FV code is usually done by a Domain
Decomposition approach. Each process is in charge of a portion of the
initial domain and ghost cells are used in order to ensure efficiently
communications between subdomains. At the border of a subdomain, faces
are duplicated, but each cell belongs to only one subdomain.
Figure \ref{ce} illustrates this spatial decomposition with 2
subdomains: for the red subdomain, the light red part corresponds to
the border cells, the dark red part to inner cells, the light green
cells are the ghost cells of this red subdomain, and finally purple
faces are the duplicated border faces.
MPI communications are only done when necessary according to the temporal level of cells that is currently computed. Border cells are tagged and are computed as soon as possible according to their temporal level. MPI asynchronous communications are used to ensure a good computation-communication overlapping.

The second level of parallelism is achieved in shared memory by using OpenMP directives (OMP DO) applied to loops concerning the cells and faces inside a subdomain.

The main problem of this spatial domain decomposition is the fact that the cells have not the same computational cost which is determined by their temporal level. To ensure a better load balancing, we give a weight to each cell of the subdomain according to its temporal level. However, the temporal level of a cell can change between iterations and  a recomputation of a new domain decomposition is needed periodically.
Despite that, the way the time is integrated leads to an important time wasted in synchronizations. The time integration implies a strict order for the cells to be processed depending on their temporal level: neighbor cells must be at the same time during computation. 
This temporal  locality information is partially lost with the current parallelization. 

\begin{figure}[h]
  \centering
  \includegraphics[width=0.98\textwidth]{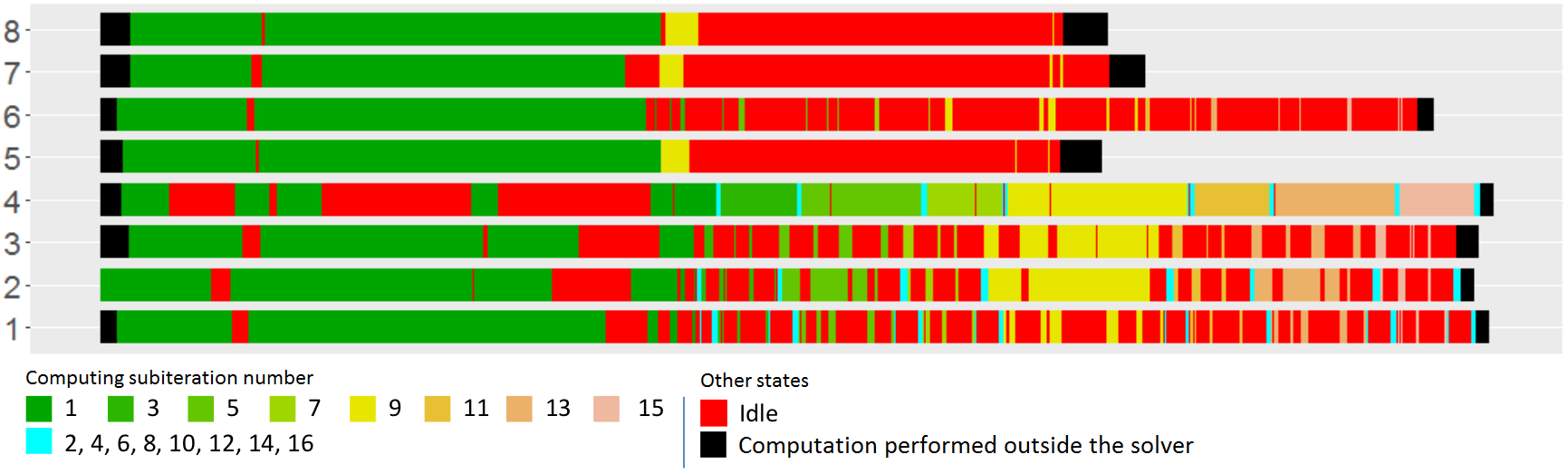}
  \caption{MPI trace for 8 processes ($\theta=4$; 16 subiterations).
  Idle time lost because of the synchronizations is colored in red.}
  \label{tl4-n8}
\end{figure}

Figure \ref{tl4-n8} shows an execution trace of the MPI version using
8 processes for $\theta=4$ (16 subiterations).
We performed an instrumentation of the code allowing to identify each
subiteration; in this trace, each subiteration is colored with a different
color, the first subiteration being colored in green.
When a subiteration concerns only $\tau_0$ cells (one over two), it is
colored in cyan.
Time spent in MPI functions and due to synchronizations is identified in red.
Computation performed outside the solver is colored in black. After the last
computation indicated in black, the process has no more work to accomplish for
the current iteration. 

Process 4 conditions the time of the iteration, being the one that
finishes last. Most of the cells of a temporal level inferior to 4 are
computed by this process 4. At the first subiteration, the only one
during which $\tau_4$ cells are computed, a lot of time is wasted in
synchronization. For other subiterations, this process 4 computes
without any interruption. 
Another interesting point is that processes 5, 7 and 8 can start the
9th subiteration (in yellow) before the others. Cells processed during
this subiteration have $\tau_3$,$\tau_2$, $\tau_1$ and $\tau_0$
temporal levels. This means that $\tau_3$ cells of those processes are
not related to the one of the other domains (they do not share
borders). Other processes (1, 2, 3, 4) can compute this subiteration lately.

This synchronization issue is one of the key elements to justify the
development of a task-based version of the aerodynamic solver. By
working on subdomains inside each process, we want to be able to
capture all the dependencies during computations and to exploit more
asynchronism with the help of a runtime system.

\section{The StarPU task-based runtime system}
\label{starpu}

There exist different libraries and frameworks to exploit task-based parallelism: e.g. SMPSs \cite{ompss}, StarPU \cite{AugThiNamWac11CCPE} , PaRSEC \cite{bosilca_parsec:_2013}, CnC \cite{budimlic_concurrent_2010}, Legion \cite{bauer2014legion}, SuperMatrix \cite{chan_supermatrix_2007}.

As said in the introduction, the main goal of task-based programming is to ensure performance portability on heterogeneous manycore distributed platforms. In this framework, the algorithm is described as a sequential task flow with data dependencies expressed through read/write attributes for each task parameter. This task flow is translated into a Direct Acyclic Graph (DAG) of tasks: the nodes represent the tasks and the edges between the nodes are the dependencies.
Then, a runtime system is in charge of scheduling the tasks over the computational resources (CPU, GPU) and of managing the data transfers.

Scientific applications are good candidates to be parallelized with such approach : they often exhibit coarse grain parallelism
on sufficiently large pieces of data. It is therefore possible to generate an interesting DAG which will allow concurrent executions
of tasks and the volume of computations for the tasks is generally sufficient to overcome the overhead of the runtime (granularity issue).
Good results have been achieved by this approach for dense linear algebra  \cite{bosilca_flexible_2011} and sparse linear algebra \cite{lacoste_taking_2014}.
There are some other works about parallelization of applications over runtimes: S3D over Legion \cite{bauer2014legion}, ScalFMM over StarPU \cite{agullo2014task}.

To construct the DAG, there are two common approaches. First, a Parametrized Task Graph (PTG) can be used \cite{cosnard1995automatic}: tasks are not enumerated but parametrized and dependencies between tasks are explicit. Another way is to use the Sequential Task Flow (STF) model. With STF, tasks are inserted from the main program and the dependencies are computed at task insertion according to data accesses \cite{allen2002optimizing}. One advantage of the later model is the fact that tasks can be inserted according to the results of previous computations.\\

The StarPU runtime system \cite{AugThiNamWac11CCPE} relies on the STF model and computation resources (e.g. CPU, GPU) are seen as workers.

Tasks are inserted from the main program through calls to {\tt starpu\_insert\_task}. The dependencies between the tasks are then computed by the runtime system according to the data accesses and how they are accessed (read, write, read-write).
This implies that the DAG is unrolled during the execution. 
The task insertion is asynchronous: the computation can start even if not all the tasks have been inserted. 
When a task is ready (i.e. all its dependencies have been fulfilled), it becomes available to the scheduler.
Then, according to the scheduling strategy, workers select tasks and execute them. Some hints are available to the scheduler: at task insertion, it is possible to give a priority to a task and some schedulers can take advantage of this information. It is also possible to have performance models which will compute a weight for a task according to various parameters (e.g. size of the data, targeted architectures) in order to help the scheduler.

\begin{figure}[!h]
  \centering
  \includegraphics[width=0.88\textwidth]{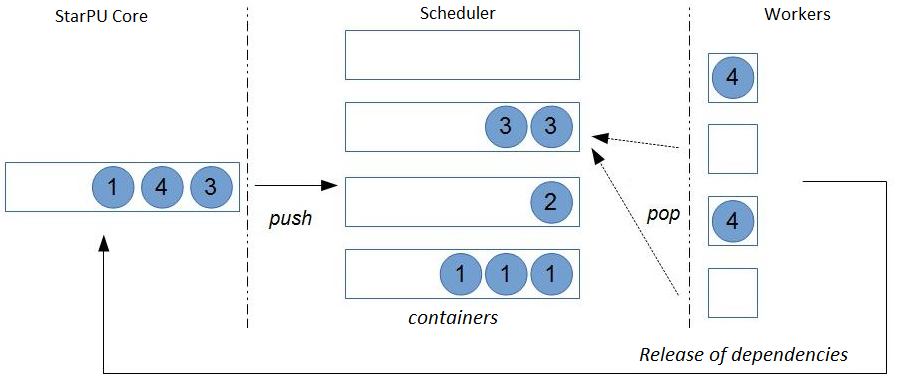}
  \caption{Scheduler with priorities (\textit{"prio"} strategy).}  
  \label{sched_prio}
\end{figure}

In StarPU, the scheduler is a modular component : it is possible to
define multiple scheduling strategies. In order to do so, the
programmer has to define containers for the tasks, a push operation
that will put ready tasks inside the right container and a pop
operation which will allow workers to choose a task to be executed.
Figure \ref{sched_prio} shows how the built-in scheduler ``prio"
works. Circles represent tasks, the number inside represents the
priority associated.
Tasks are inserted with a priority from 0 to 4. When they are ready,
they are pushed to the right container.
The containers are FIFO lists : one list exists for each allowed priority.
For the pop operation, the workers will check every list by order of
decreasing priority until they find a task to execute.

\subsection{Parallel tasks and worker-contexts}

It is possible to exploit existing OpenMP code within StarPU through the use of context \cite{cojean:hal-01181135}. A context can be seen as a set of computational resources with a scheduling strategy in which tasks can be submitted. Contexts can be nested.

When no scheduling strategy is specified, the context is seen as a worker (called worker-context): when a task is pushed on this worker-context, it is supposed to be executed on all its computational resources. 

To use an OpenMP code inside a worker-context, a specific initialization task that will bind OpenMP threads to the CPUs of the worker-context must be used. 
By default, tasks are inserted in a global context. So, if one wants to create 4 worker-contexts of 4 CPUs, he has to create a main-context with a scheduling strategy that will contain these worker-contexts and insert tasks in this main-context.

\subsection{Distributed parallelism with StarPU}

\begin{figure}[!ht]
  \centering
  \subfloat[Implicit Communications]
  {
  \includegraphics[width=0.42\textwidth]{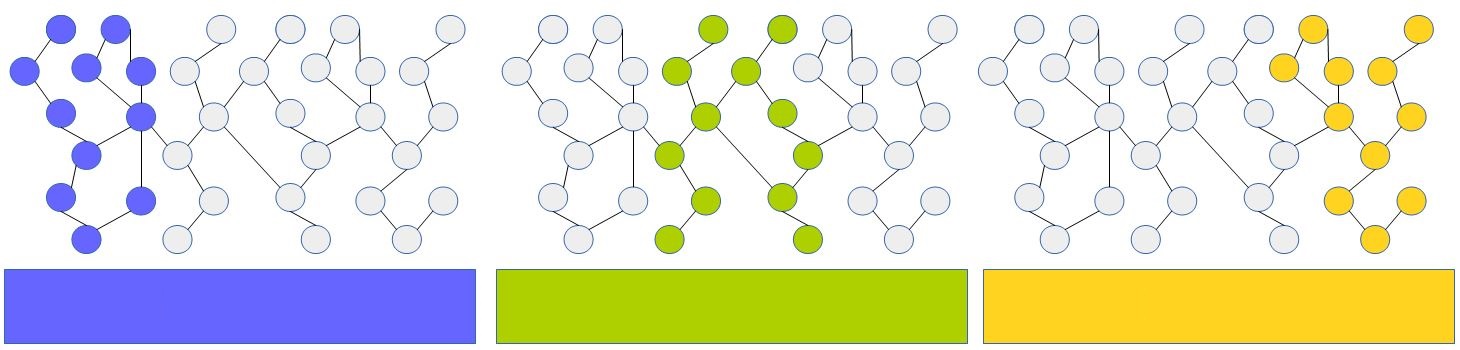}
  }
  \subfloat[Explicit Communications]
  {
  \includegraphics[width=0.42\textwidth]{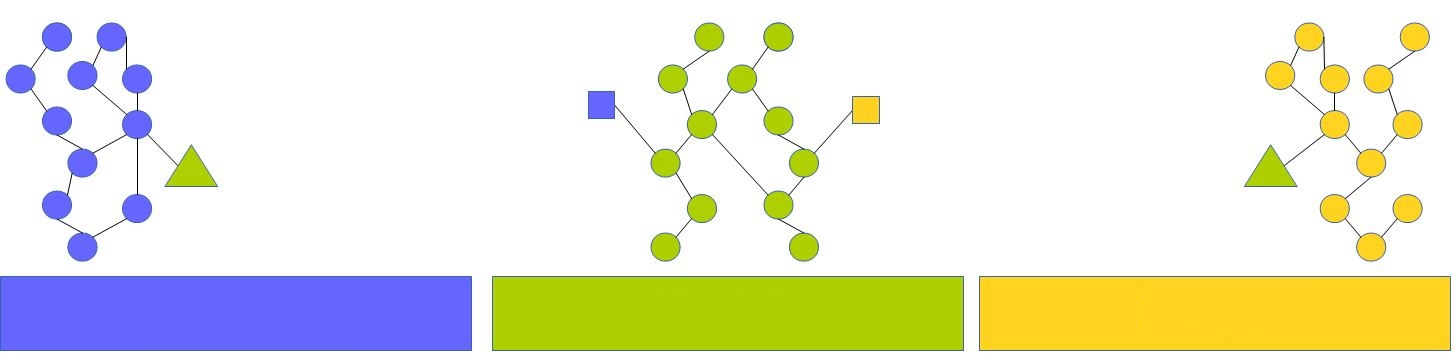}
  }
  \caption{DAG representation for three nodes and using implicit or explicit communications.}
  \label{dag_mpi}
\end{figure}

There are two ways to use MPI parallelism with StarPU. First, StarPU provides {\tt starpu\_mpi\_insert\_task} that lets StarPU handle all the communications and data transfers. With this approach, all the tasks must be inserted in all the nodes of the cluster. Communications are implicit and each node knows which node will execute a particular task.

{\tt starpu\_mpi\_isend\_detached} 	and {\tt starpu\_mpi\_irecv\_detached} allow to describe communications explicitly. When using those primitives, communications are consistent with the computation dependencies inferred by task insertion inside the nodes.

Figure \ref{dag_mpi} shows the difference between the two approaches while using three nodes. The same DAG is considered. In the implicit case, the whole DAG is present in each node and only a portion of it is useful. In the explicit case, each node only contains the tasks it will execute and communication tasks are present.

\section{MPI + Task design of the aerodynamic solver}
\label{impl}

\subsection{Computation Elements}

As the first level of parallelism we can exploit in FLUSEPA is spatial, the natural way to described tasks is to use multiple subdomains. In the DAG, we want to express dependencies in such a way that allows the maximum concurrency. In order to achieve this, we use an algorithmic abstraction called ``Computation Element'' (CE).

The code mainly manipulates faces and cells and this must be exploited to have a good task parallelization. Four main basic computation patterns can be found in the original code: computation on cells, computation on faces, computation on faces using cell values, computation on cells using face values. However, the way the code has been written does not lead to a clear distinction between those patterns when they were used, and thus a rewriting step has been performed. Most of the computation kernels of the aerodynamic solver will now correspond to one of these computation patterns in order to achieve a well-structured task version (cf. Algorithm \ref{task_insertion}).

\begin{figure}[!h]
  \centering
  \includegraphics[width=0.50\textwidth]{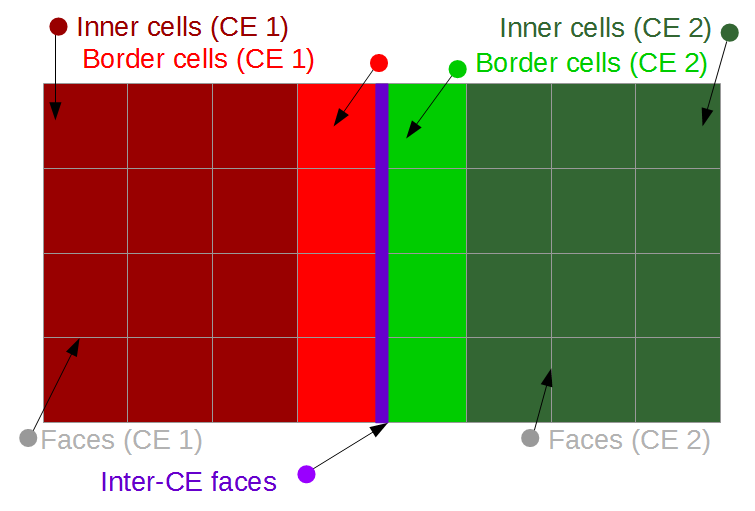}
  \caption{Illustration of every component for two neighboring CEs.}  
  \label{ce}
\end{figure}

The domain partitioning is done with the graph partitioner SCOTCH\footnote{https://gforge.inria.fr/projects/scotch/} \cite{pellegrini1996scotch}. For each subdomain created, we associate a CE containing informations to retrieve each field or flow value (See Section \ref{fvas}). The different components of a CE are shown at Figure \ref{ce}: there are two CEs, a red one and a green one.
Border cells (light red and light green) and inner cells (dark red and dark green) of the CEs are distinguished. Between CEs, we also have inter-CE faces (purple). Because we have a unstructured mesh, all these topological informations are precomputed for each CE after the partitioning step.

\subsection{Task generation}

\begin{algorithm}
\caption{Task insertion for the Aerodynamic Solver}
\begin{algorithmic}[1]
\STATE Insert tasks for time step computation
\STATE Insert tasks for classification of cells in temporal levels
\STATE Wait all tasks
\STATE Compute hints (scheduling and packing algorithm)
\STATE \textit{Temporal adaptive loop:}

\FOR{subiteration=1 \TO $2^\theta+1$}
	\STATE Compute $\tau$
	\STATE Insert task for predictor (0 to $\tau$)
	\FOR{$\tau'=\tau$ \TO $0$}
		\STATE Insert task for corrector ($\tau$)
	\ENDFOR
\ENDFOR	 
\STATE Wait all tasks
\STATE Send informations to master process
\STATE Foreach cells update intensive values
\STATE Wait all tasks
\end{algorithmic}
\label{task_insertion}
\end{algorithm}

In order to generate tasks, we mainly use the different data types we described previously and exploit the associated computation patterns. We rely on the STF model of StarPU.
For each pattern, we used a ``foreach" function that is in charge of generating the right tasks. We mainly exploit Algorithm \ref{ta_algo}, each line being converted in one or several ``foreach''.

The task insertion follows Algorithm \ref{task_insertion}. The time step computation (line 1) and the classification of cells in temporal levels for each CE (line 2) are also done using tasks. Currently, we have one synchronization after the classification of cells in temporal levels. In order to know which task must be inserted, the temporal levels inside each CE must be known. Some CEs do not contain cells of some temporal levels and this fact is exploited to avoid insertion of useless empty tasks.

The foreach functions can generate a high number of tasks and in particular chains (succession of tasks that depend only of one previous task). In order to reduce the number of tasks, we implemented a strategy to pack tasks at runtime. 
A pack is then a single task composed of multiple elementary tasks.

Computation kernels are written in such a way that they only write one component of a CE and the pack mechanism relies on that. We have 3 kinds of task packs, each one for modifying data either on border cells, inner cells or faces.
For each CE, these task packs exist. 

The foreach functions know which type of component (face or cell) will be modified. If it is the same at the previous one, tasks will be added to the corresponding pack associated to the CE. Otherwise, the previous packs are inserted as tasks and new packs are created.

Another case that generates chains of tasks happens when cells of a given temporal level are present only in the inner cell component of a CE. For each CE, we check if it is the case at line 4 of Algorithm \ref{task_insertion}.
To handle this particular context, we introduce a 4th kind of pack called ``large task pack". As there is no interaction with other CEs, they contains all the tasks for a given temporal level and a given CE, disregarding the type of component currently modified.

Optimizing this point is critical because tasks that work on cells with low temporal levels are numerous and they work on a low number of cells.
This fact is shown in Table \ref{ta_table} which gives the proportion of cells in each temporal level and the proportion of associated computation cost for the test case we use in Section \ref{xp} when fixing the maximum temporal level to 2.
 The large majority of cells are of the higher temporal level ($\tau_2$). Few cells belong to low temporal levels, so we can expect that most of them won't be part of the border component of a CE.

\begin{table}[!h]
\centering
\begin{tabular}{| c | c | c | c | c | r |}
\hline
& $\tau_0$ & $\tau_1$ & $\tau_2$ \\
\hline
\small{Cells} & \small{0.05\%} & \small{2.42\%} & \small{97.53\%} \\
\hline
\small{Computation} & \small{0.20\%} & \small{4.72\%} & \small{95.08\%} \\
\hline
\end{tabular}
  \caption{Cell distribution and associated computation cost ($\theta=2$).}
  \label{ta_table}
\end{table}

\begin{figure*}[!ht]
  \includegraphics[width=1.01\textwidth]{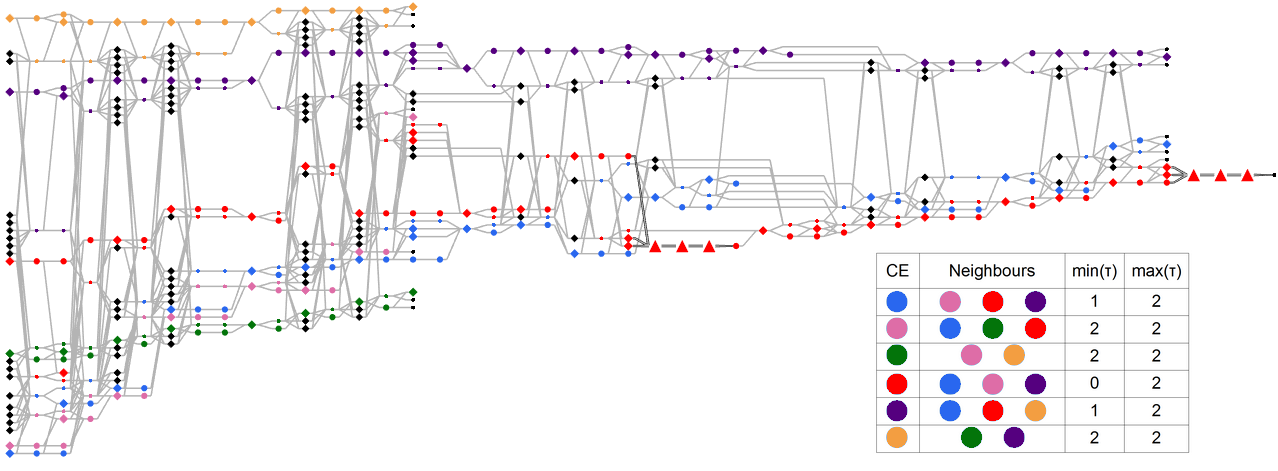}
  \caption{DAG for a computation with 6 CEs and $\theta=2$.}  
  \label{dag}
\end{figure*}

Figure \ref{dag} shows a DAG for an iteration with a maximum temporal level $\theta=2$, three temporal levels of cells, 4 sub-iterations and 6 CEs. Colors represent the different CEs. The red CE contains cells of the three levels, two other CEs (blue and purple) contain $\tau_1$ and $\tau_2$ temporal level cells. The remaining three CEs contain only $\tau_2$ temporal level cells.

The colored diamonds correspond to computations that modify data for faces of a CE, black diamonds for inter-CE faces, small circles for border cells and large circles for inner cells. Triangles represent large task packs: in this case, each large task pack contains more than 10 subtasks. 
In this DAG, we can notice that $\tau_0$ temporal level cells are only present in the inner component of the red CE as there are large task packs.

Increasing the number of CEs would enlarge the width of the DAG and using more temporal levels would enlarge its height for some CEs. The DAG can really be unbalanced according to the cell distribution in temporal levels inside each CE, so the way the graph is traversed can have a strong impact on the efficiency of the computation.

\subsection{MPI + Task parallelization}

As we insert tasks according to previous computations, we decided to rely on explicit communications. 

Each process gets a domain after the domain partitioning. At this moment, for each process, border cells are identified and border faces are duplicated. In order to distinguish border cells of the domain associated to a process and border cells of the component of a CE, we note the first ones ``MPI-border cells".

\begin{figure*}
  \includegraphics[width=0.95\textwidth]{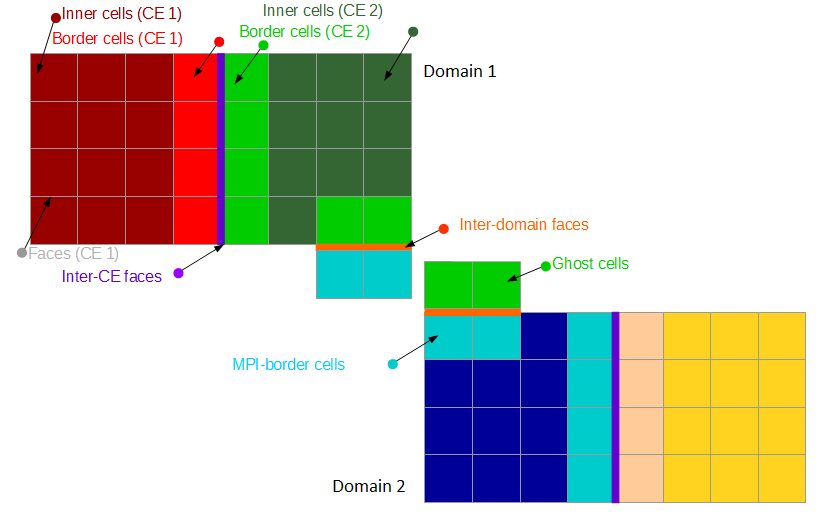}
  \caption{Two neighboring domains, each one with two neighboring CEs.
  The components introduced for the distributed version are inter-domain faces, ghost cells and MPI-border cells.}  
  \label{ce_mpi}
\end{figure*}

These MPI-border cells are then included to the border component of the CE they belong. After that, each process communicates with its neighbors to complete informations for the common border cells. For each (local CE, foreign CE) couple, we create a border ghost cell component.

Figure \ref{ce_mpi} shows two domains with two CEs each. For the two CEs that share a border, we can see the new components : MPI-border cells, inter-domain faces and ghost cells.

For the management of communications, a task copies border cell content to a temporary buffer and this buffer is sent using {\tt starpu\_isend\_detached}. For receiving data, the {\tt starpu\_irecv\_detached} operation is used in a temporary buffer and then, we copy the data in the ghost cell component.

We use a temporary buffer because we do not want to send the whole border component of a CE: some border cells are not MPI-border cells and when processing low temporal levels, a fraction of the MPI-border cells is concerned.

Before foreach functions that uses cells in read mode, we insert communications as tasks.

Our task-based implementation is able to exploit 3 levels of parallelism : between computation nodes, MPI is used; inside a SMP node, StarPU tasks are used; and tasks can be parallel by using OpenMP.

\subsection{Parallelization effort}

In order to achieve a satisfactory parallel task-based version of the aerodynamic solver,
we first designed a symbolic version that mimics the behavior of the solver, but without
performing the actual computations. 
This way, we have been able to try different patterns for the computation elements leading
to different amounts of computation and different data accesses. 
This prototype has been very helpful and allowed us to follow a precise methodology to refactore
the original code. This refactoring step implied to split some original subroutines and 
modify their signatures.

Moreover, we relied on a functionality that allowed us to execute some parts of the code from
the original version and others using the new task-based version. This was possible through
the use of the {\tt starpu\_pause} and {\tt starpu\_resume} primitives.
We add a barrier where the runtime 
waits for all inserted tasks to complete;
 then, some computations
can be done by using some subroutines of the original version of the solver, and finally the runtime
is restarted and tasks are inserted again.
Of course, this implies additional synchronizations but this way, it was possible to build the task-based version
step by step with the possibility to make non-regression tests to ensure the correctness of the computations
with regards to the initial version.

When adding parallel OpenMP tasks, no additional modification was required in the kernels since the original
code already had OpenMP pragmas.
Concerning MPI communications, we used the scheme from the original code. Each time we had a {\tt MPI\_Waitall}
in the original code, we inserted an explicit communication using dedicated tasks. 
With this task-based version, there is no need to identify the best moment to send the data since the runtime
will be able to determine it according to the dependencies in the DAG associated to the solver.

\section{Experimental study}
\label{xp}
For the experimental validation of this work, we used a cluster with
nodes composed of two 8-cores Sandy Bridge with 64 GB of RAM.
Our current implementation is dedicated to distributed clusters with
multicore SMP nodes, so no accelerator is used for the moment.

\begin{figure}[!ht]
  \centering
  \subfloat[Geometry of the launch pad]
  {
  \includegraphics[width=0.30\textwidth]{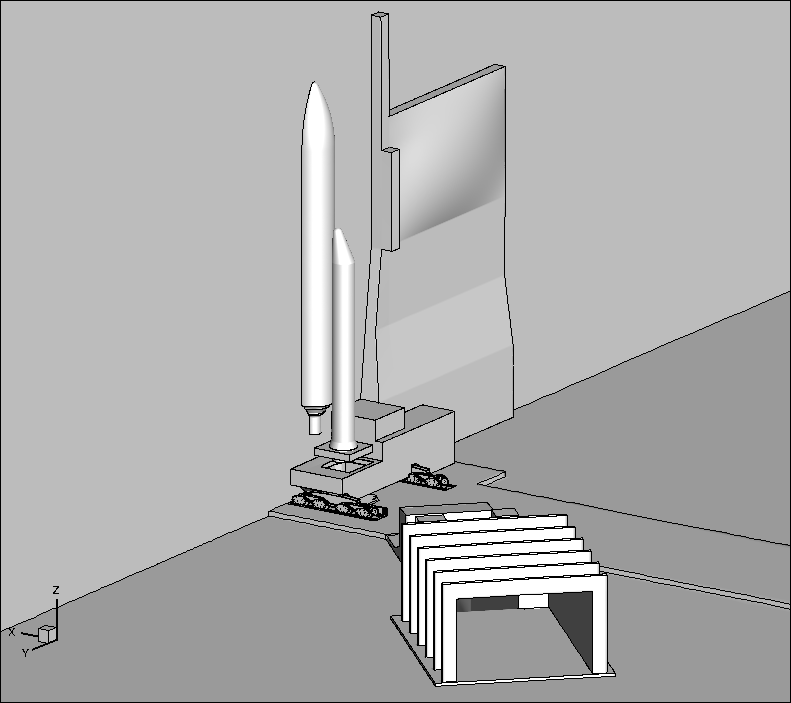}
  }  
  \subfloat[Top view of the mesh]
  {
  \includegraphics[width=0.30\textwidth]{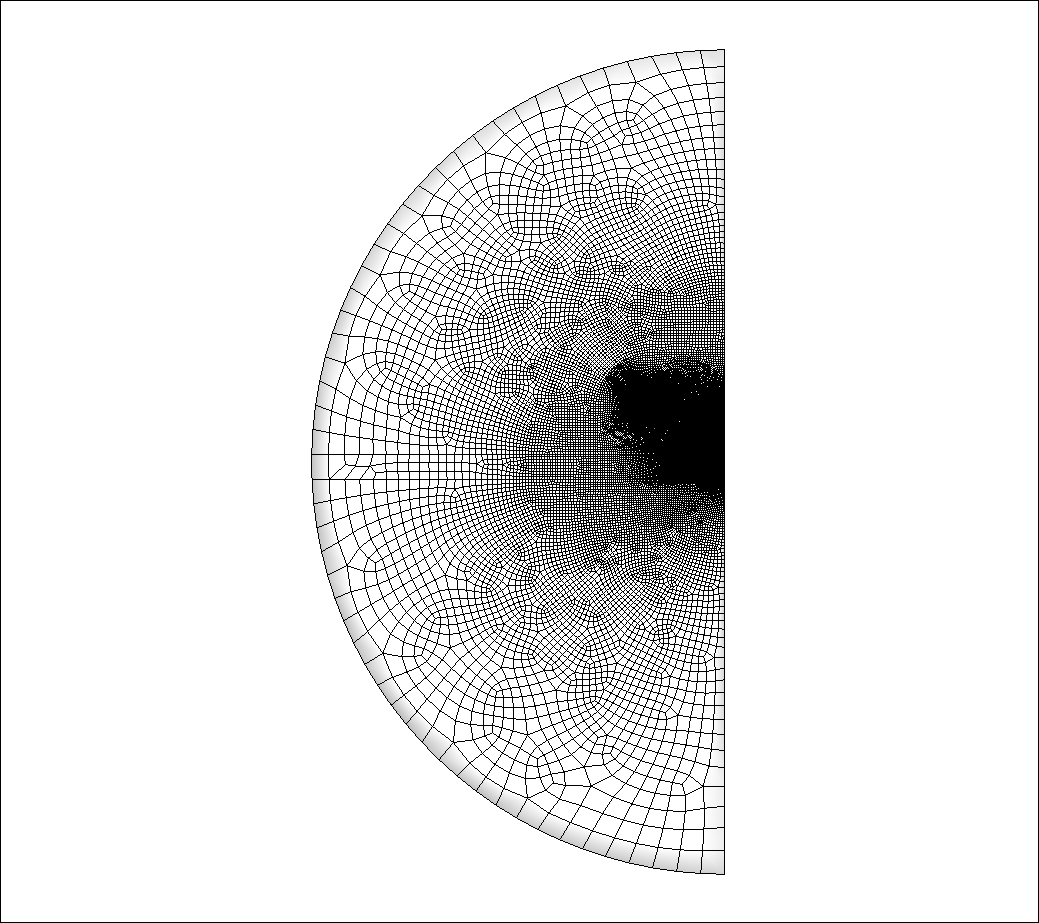}
  }
  \subfloat[Side view of the mesh]
  {
  \includegraphics[width=0.30\textwidth]{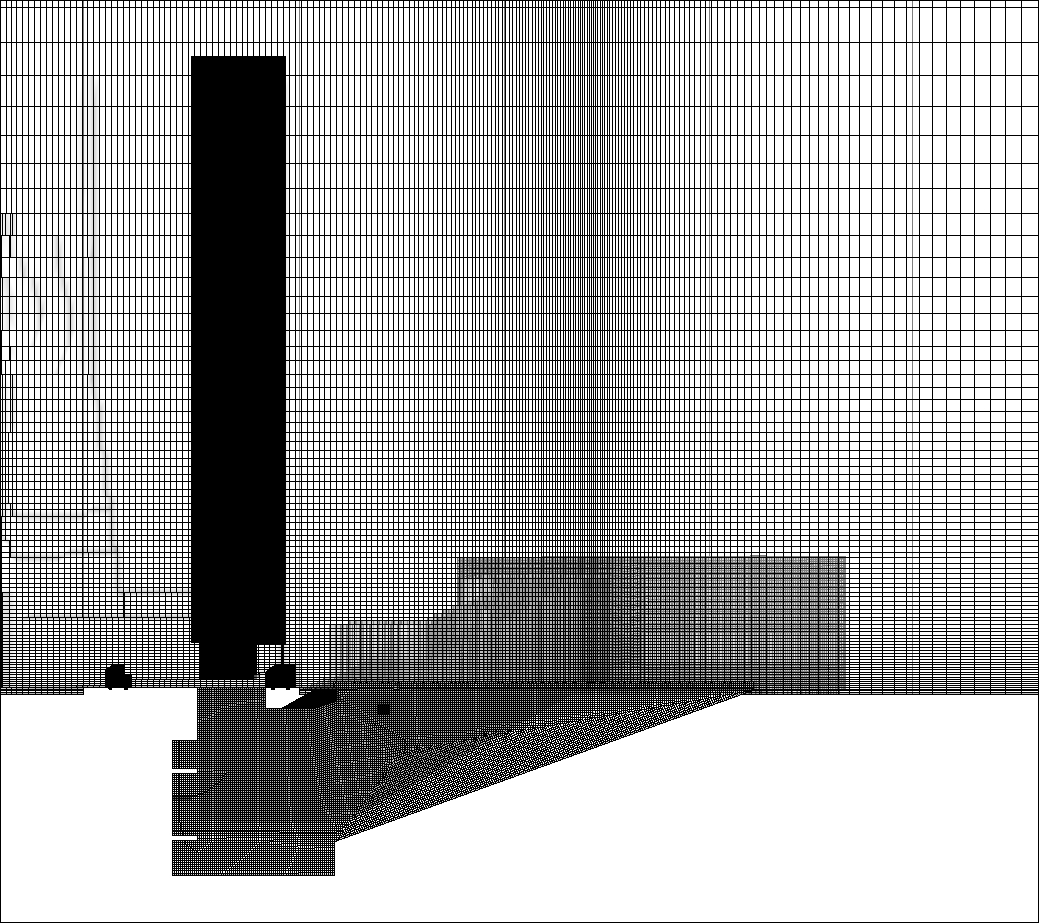}
  }
  \caption{Mesh for the take-off blast wave computation.}
  \label{mesh}
\end{figure}

\begin{figure*}[!ht]
  \centering
  \includegraphics[width=.19\linewidth]{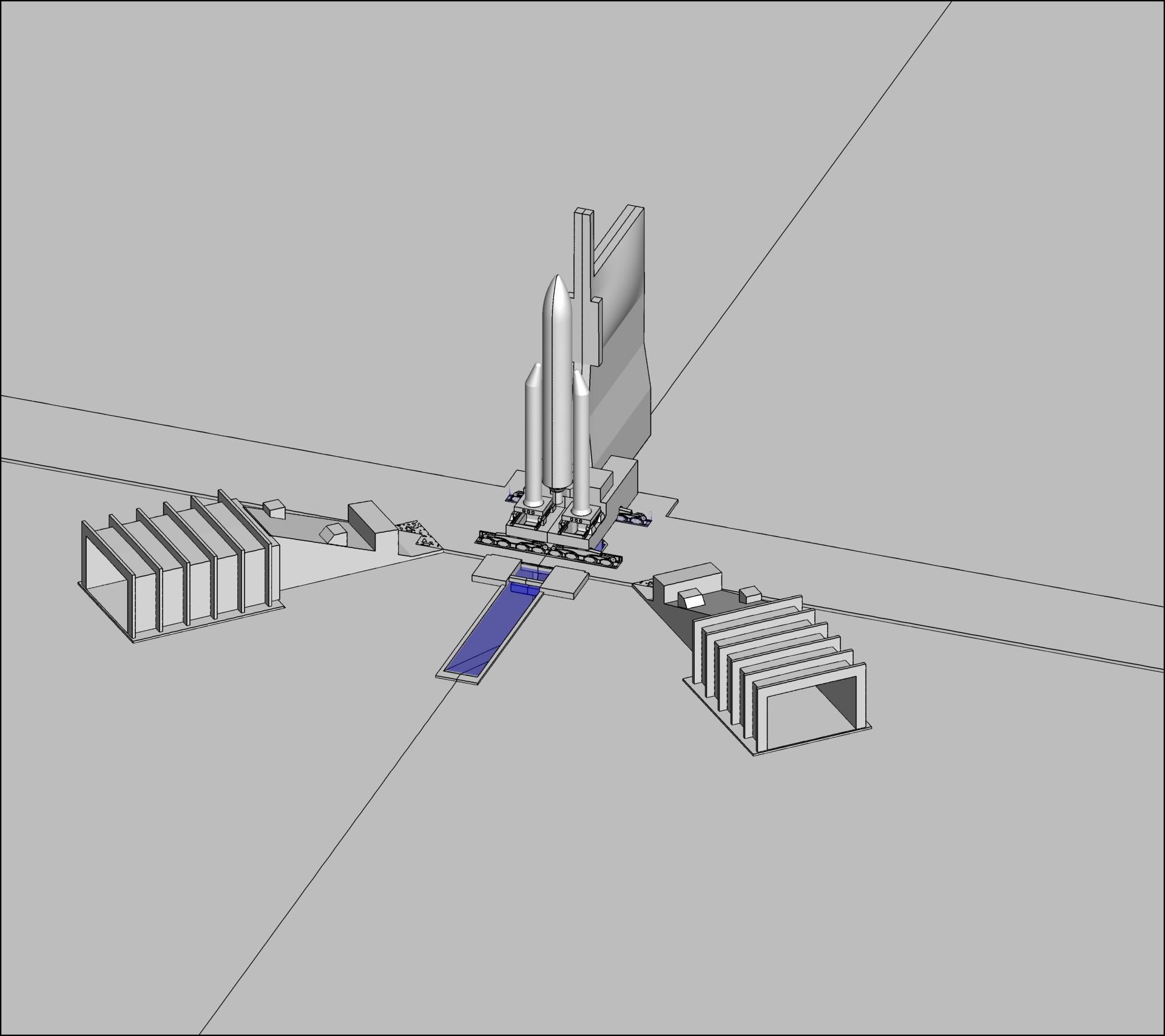}
  \includegraphics[width=.19\linewidth]{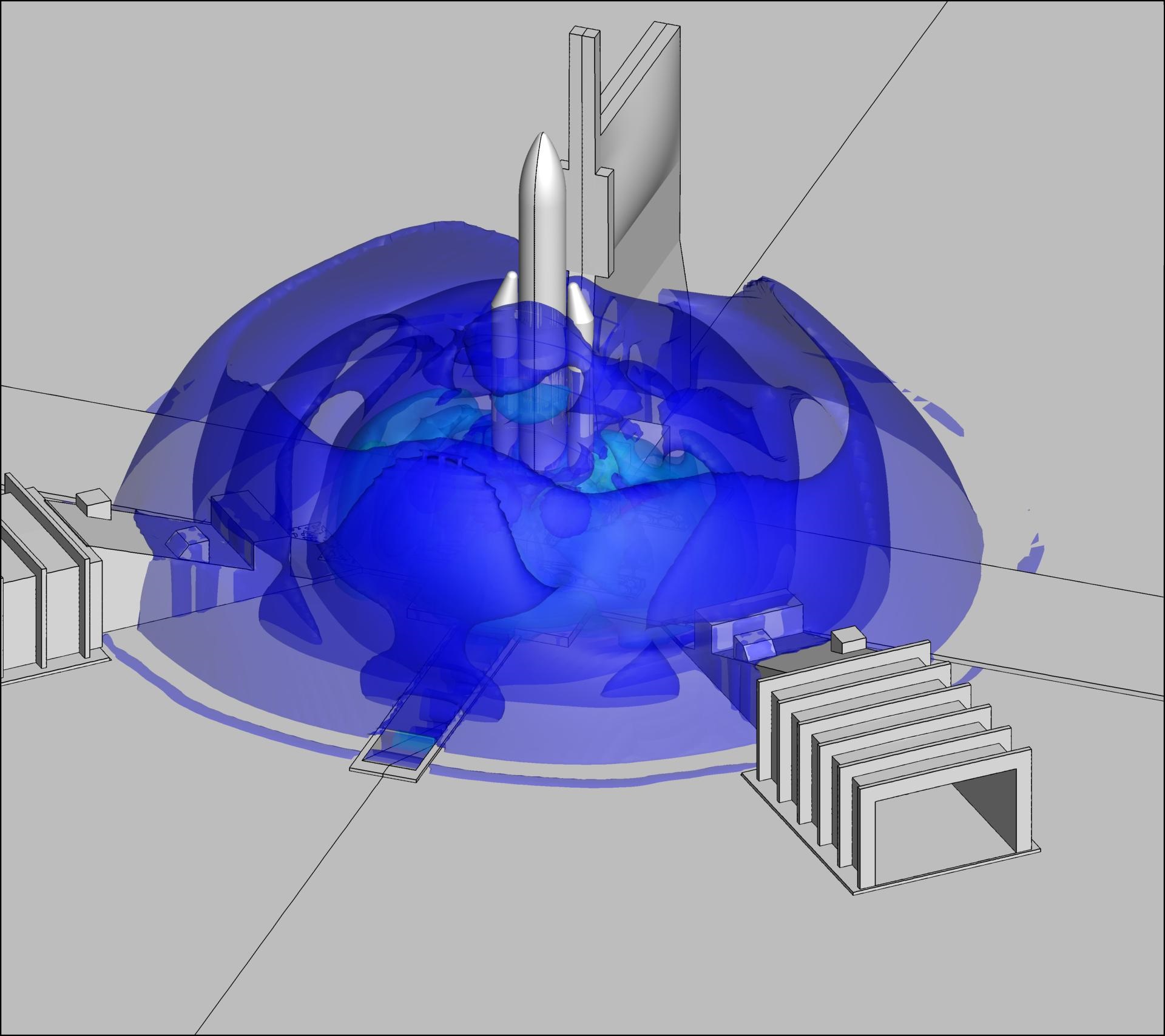}
  \includegraphics[width=.19\linewidth]{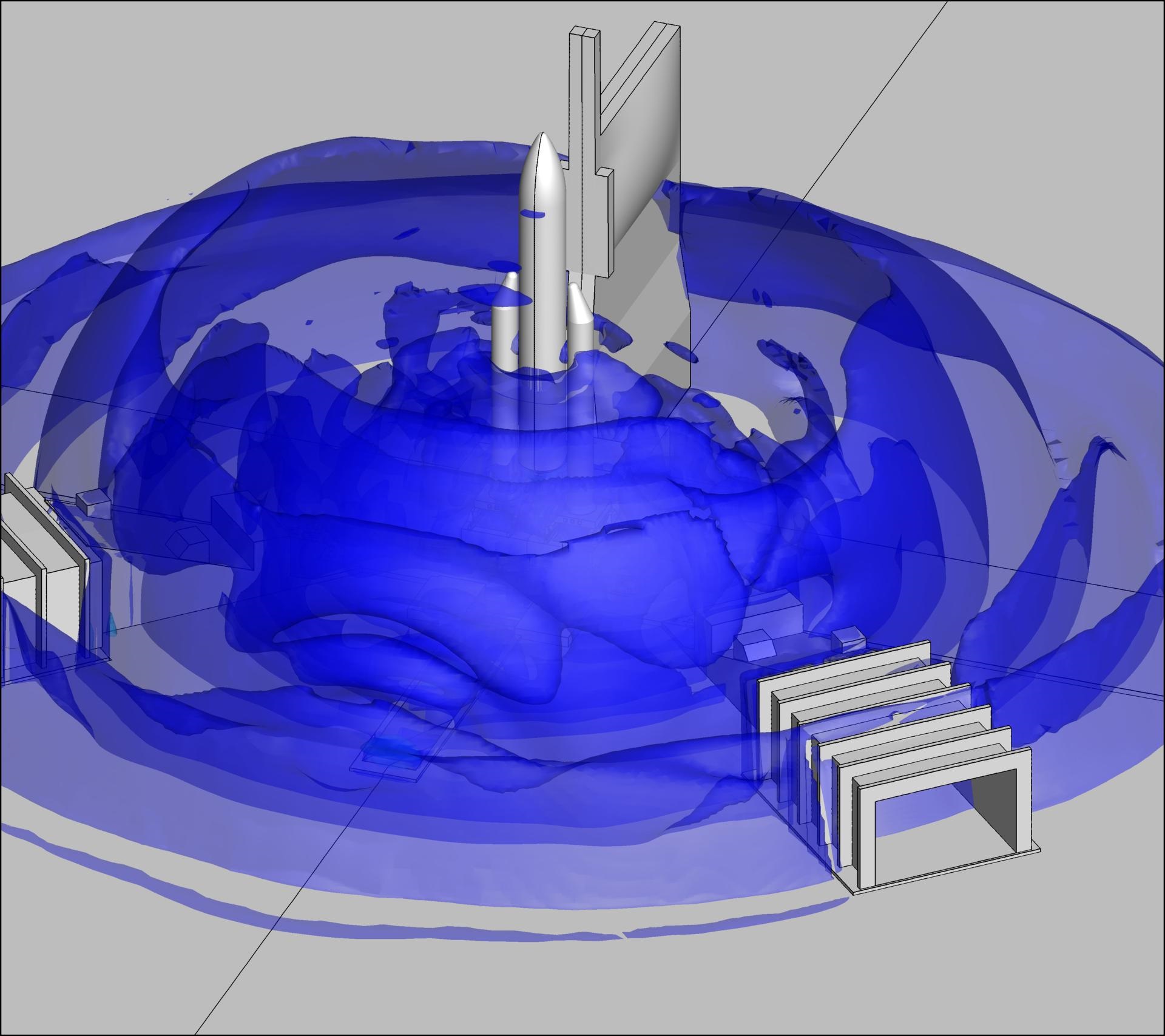}
  \includegraphics[width=.19\linewidth]{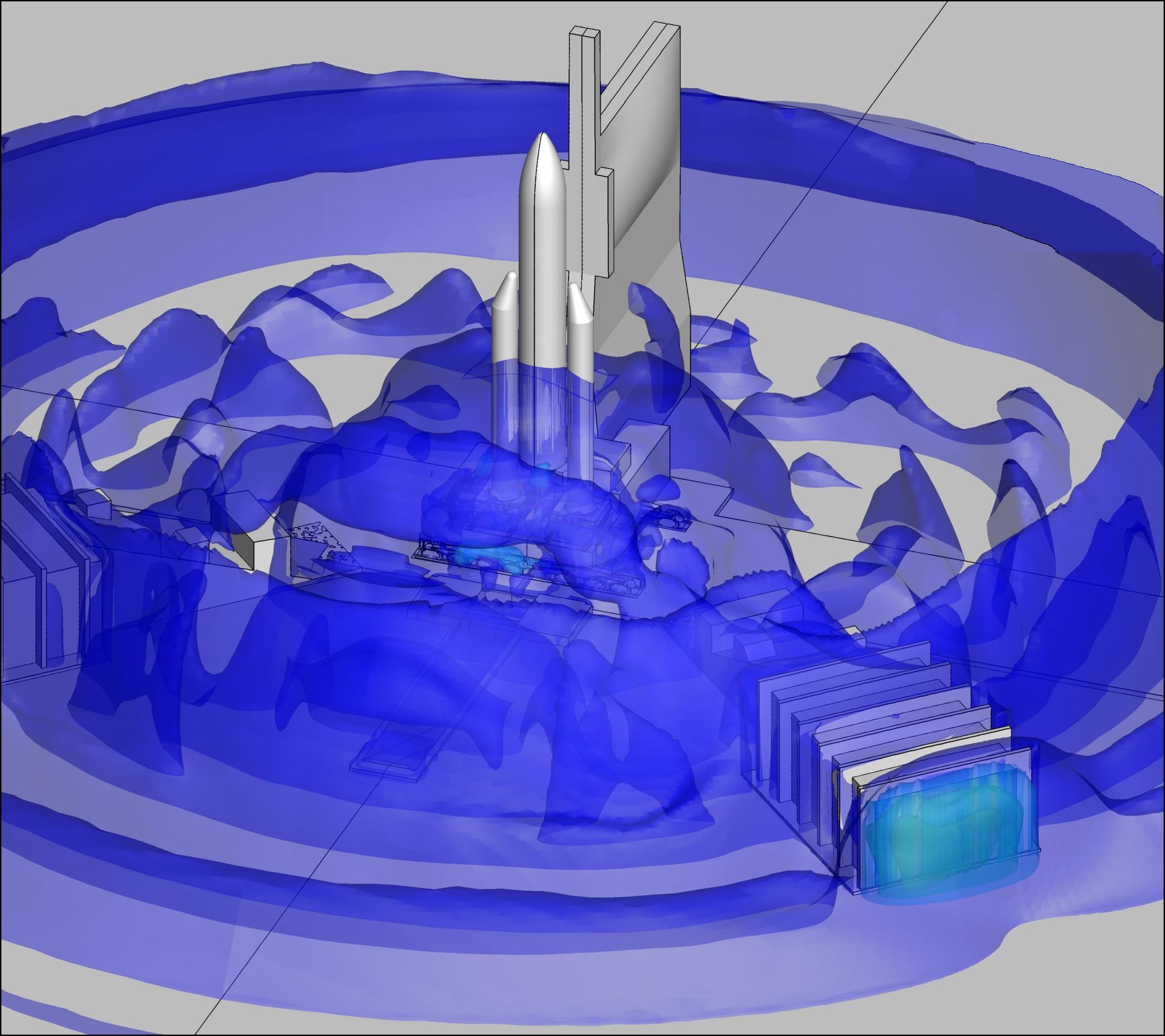}
  \includegraphics[width=.19\linewidth]{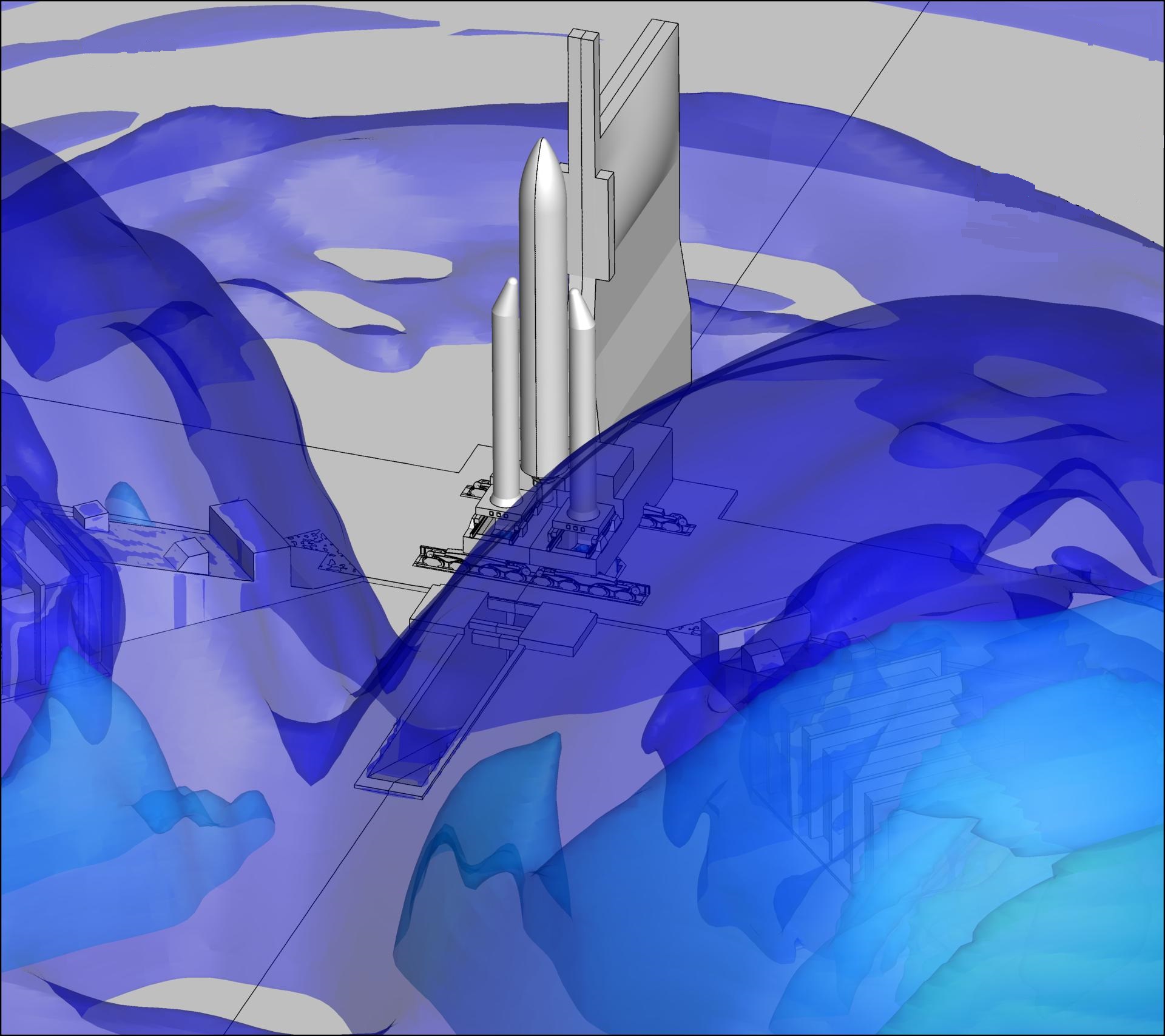}  
  \caption{Take-off blast wave computation.}
  \label{ods}
\end{figure*}

The test case from Airbus Defence \& Space is the computation of a
take-off blast wave. For an Ariane 5 rocket, boosters provide 90\% of
the thrust at lift-off and the objective is to compute the resulting
blast-wave implied by the ignition of the boosters.
The mesh is composed of 10M cells and is finer around area of interest
(Figure \ref{mesh}). During take-off blast wave propagation, there are
two overpressures: the first one is due to ignition of the boosters,
the second one is due to ducts. Those events are visible at Figure
\ref{ods}: the comeback of the wave can be seen from the 4th picture.

\begin{figure*}[!ht]
  \centering
  \subfloat[Proportion evolution of cells]
  {
  \includegraphics[width=0.33\textwidth]{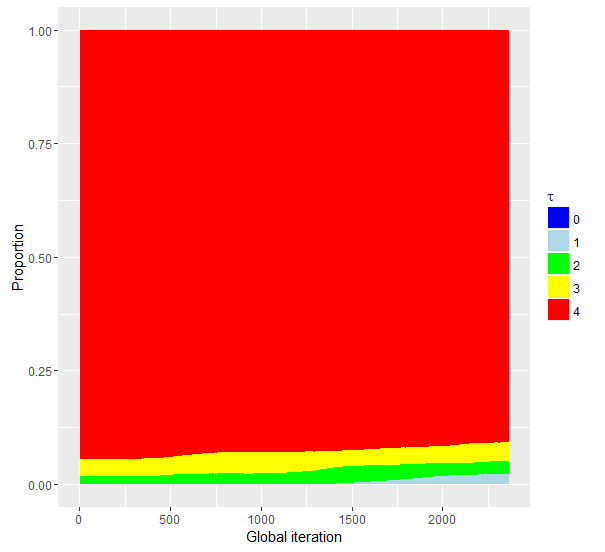}
  \label{temporal_class_evolution}
  }
  \subfloat[Proportion of computation cost]
  {
  \includegraphics[width=0.33\textwidth]{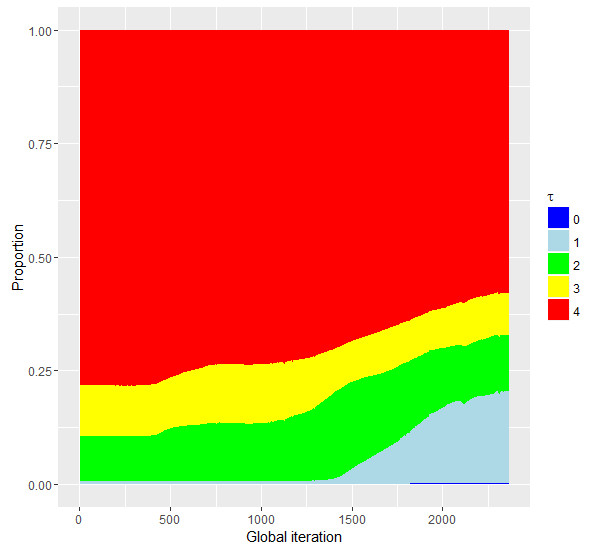}
  \label{cost_evolution_prop}
  }  
  \subfloat[Global cost]
  {
  \includegraphics[width=0.33\textwidth]{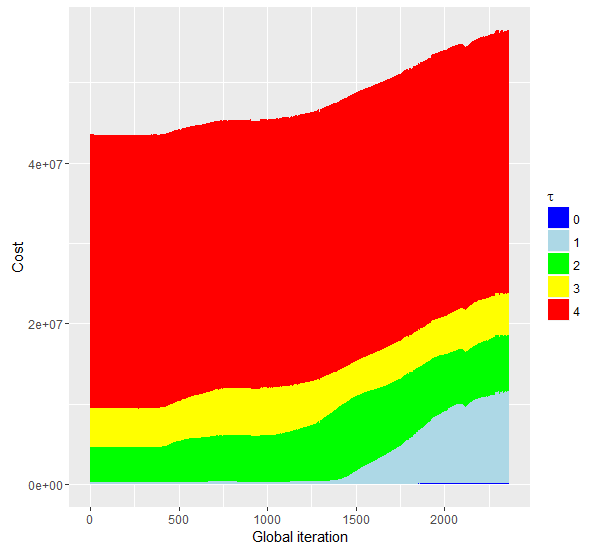}
  \label{cost_evolution_tots}
  }    
  \caption{Evolution of the number of cells and of the computation cost by
    temporal level $\tau$ and by iteration.}  
  \label{ta_iter_evolution}
\end{figure*}

We consider a case with $\theta=4$.
In this computation, the cell repartition in different temporal levels
and the theoretical computation cost associated evolve during the
computation, which is shown at Figure \ref{ta_iter_evolution} and
Tables \ref{tl_repartition_step_1_tl_4} and
\ref{tl_repartition_step_2_tl_4}.
The tables show the proportion of cells and the associated computation
cost at two different times of the computation: Table
\ref{tl_repartition_step_1_tl_4} at the start of the computation and
Table \ref{tl_repartition_step_2_tl_4} at the 2300th iteration.

Figure \ref{temporal_class_evolution} shows how the proportion of
cells in each temporal level evolves.
There are few cells of the lower temporal level $\tau_0$ and they are
not visible on the figure.
We can observe that the proportion of cells of temporal level $\tau_1$
evolves during the computation, but they represent at most 2\% of the cells.
Figure \ref{cost_evolution_prop} shows the relative cost of each level
$\tau$ and its evolution during the computation. 
Figure \ref{cost_evolution_tots} shows the global cost of an iteration
(in floating-point operations) during the computation. It shows  that
the cost of one iteration varies and that it gets more and more
expensive while the computation progresses.
The computation cost associated to $\tau_1$ cells evolves a
lot. According to Tables \ref{tl_repartition_step_1_tl_4} and
\ref{tl_repartition_step_2_tl_4}, it increases from 0.45\% to 12.45\%
between the start of the computation and the 2300th iteration.\\

For our task-based parallel implementation, we focus now on one
iteration of the aerodynamic solver and we first study in shared
memory the impact of several parameters: the priority strategy for the
scheduling of tasks, the number of CEs, the number of parallel workers.
Once we have evaluated the impact of those parameters, we will evaluate
the implementation at two different iterations with different cell
proportions in the temporal levels. Then, we will evaluate a
distributed memory version combining all the levels of parallelism.

\begin{table}[!h]
\centering
\begin{tabular}{| c | c | c | c | c | c |}
\hline
 & $\tau_0$ & $\tau_1$ & $\tau_2$ & $\tau_3$ & $\tau_4$ \\
\hline
Cells & 0.00004 \% & 0.06106 \% & 1.65942 \% & 3.66178 \% & 94.61769 \% \\
\hline
Computation & 0.0006 \% & 0.4479 \% & 6.0858 \% & 6.7147 \% & 86.7510 \% \\
\hline
\end{tabular}
  \caption{Cell distribution and associated computation cost at the
    start of the computation ($\theta=4$).} 
  \label{tl_repartition_step_1_tl_4}
\end{table}

\begin{table}[!h]
\centering
\begin{tabular}{| c | c | c | c | c | c |}
\hline
 & $\tau_0$ & $\tau_1$ & $\tau_2$ & $\tau_3$ & $\tau_4$ \\
\hline
Cells & 0.02 \% & 2.00 \% & 3.04 \% & 5.01 \% & 89.93 \% \\
\hline
Computation & 0.22 \% & 12.45 \% & 9.48 \% & 7.81 \% & 70.04 \% \\
\hline
\end{tabular}
  \caption{Cell distribution and associated computation cost at the
    2300th iteration of the computation ($\theta=4$).} 
  \label{tl_repartition_step_2_tl_4}
\end{table}

\subsection{Shared memory study}
\subsubsection{Impact of priorities and of number of CEs}

For this study, we don't use parallel workers, so each CPU is associated to only one worker.

We work with the StarPU built-in ``prio'' scheduler. In this scheduler, tasks are pushed in different priority queues and they are popped by priority order. 

We consider two strategies. In the first one, we don't give any priority to any CE. For the second one, we give to tasks a priority according to the CE they belong to. We want to give an advantage to the CEs with low temporal levels. However, when giving a priority to a task in StarPU, it doesn't automatically propagate the priority to the predecessors of the task in the DAG. So, it is also necessary to give a good priority to the predecessors of the tasks.

To achieve this goal, we give the highest priority to CEs which contain cells of temporal levels 0 or 1. Then for each CE, we compute a priority : we evaluate the lower distance from this CE to the other CEs previously prioritized. Then we map priorities according to the distance computed: the lower the distance, the higher the priority.

\begin{figure}[!h]
  \centering
  \includegraphics[width=0.49\textwidth]{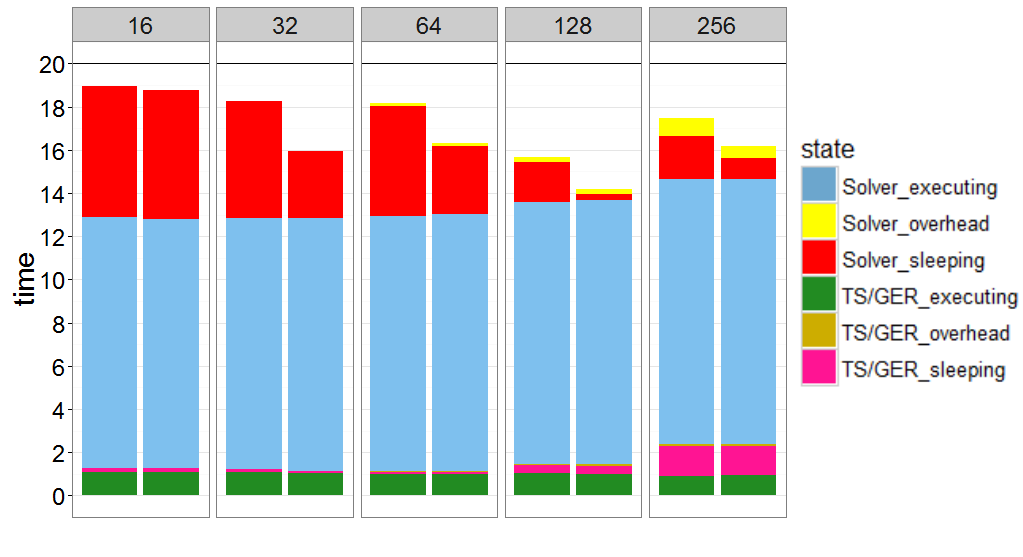}
  \caption{Impact of priority strategy with a varying number of CEs.}
  \label{sm-no-par}
\end{figure}

The result for our test case are shown at Figure \ref{sm-no-par}. The reference time (20.02s) is the one for a computation with only 1 CE and one unique parallel worker which uses the 16 cores of the node; this is the configuration that mimics the best the previous OpenMP version in shared memory. The first observation is that this reference configuration with 1 CE is the worst in terms of performance. For a number of CEs from 16 to 256, we compare the elapsed time with our priority strategy (on the right) and without priority (on the left).

We show the state of the different workers during one iteration. The global size of the bar indicates the time (in seconds) needed to complete one iteration and the fill colors correspond to the proportions spent in each state (executing, sleeping, overhead).
With the {\tt starpu\_profiling\_worker\_get\_info} function, it is possible to recover executing and sleeping time for each worker during a measurement interval. 
The sleeping state means that the worker is ready but there is no ready task available.
We count as ``overhead'', the difference between the duration of the measurement interval and the sum of the executing and sleeping time.
The overhead state contains, among other things, time spent to compute dependencies and to insert tasks. 

The TS/GER part corresponds to lines 1 to 3 of Algorithm \ref{task_insertion}, including the ``Wait all tasks'' step. The Solver part corresponds to lines 4 to 15. 
We observe that the time spent in computation (blue and green bar) does not evolve when we increase the number of CEs, whether or not the priority strategy is used. Concerning computations from 16 to 128 CEs, sleeping state (red and dark pink bar) is reduced by decomposing in more CEs and by using the priority strategy. This strategy is beneficial as soon as we use 32 CEs. Concerning the overhead (yellow and gold bar), it evolves linearly with the number of CEs and the priority strategy has no influence. We detail below what happens for 128 and 256 CEs.

\begin{figure*}[!ht]
  \centering
  \subfloat[Execution trace without priorities (t=15.565s)]
  {
  \includegraphics[width=0.69\textwidth]{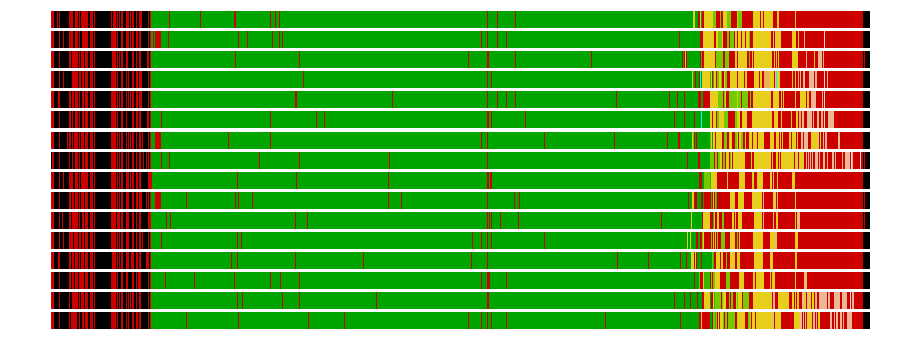}
  }
  
  \subfloat[Execution trace with priorities (t=14.184s)]
  {
  \includegraphics[width=0.69\textwidth]{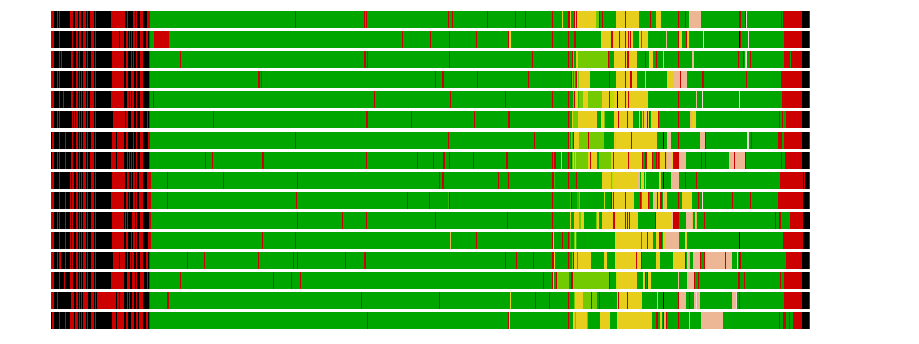}
  }
  \caption{Two execution traces involving a 128 CE computation, one without and the other with priorities. By scheduling the tasks in a different order, it is possible to reduce idle time.}
  \label{gantt1}
\end{figure*}

First, let us consider the computation with 128 CEs to highlight what happens when we use or not the priority strategy. Figure \ref{gantt1} is a Gantt Diagram where each horizontal bar represents a worker. Idle time and overhead are always in red. 

The first step of the computation is in black (TS/GER, lines 1 to 3 of Algorithm \ref{task_insertion}).
For the solver part of the computation (lines 4 to 13), tasks are colored according to the subiteration they compute. 

Of course, the same DAG is built and only the order of computations changes when using the priority strategy. In the trace without priorities, we observe that at the end of the computation, there is a large starvation zone: workers spend time in sleeping state (in red) because there is no task available. When using priorities, this starvation is much less important: we perform the DAG traversal in a way that allows more tasks to be available at the end of the computation. It is also noticeable that we managed to finish the scheduling by task concerning the first subiteration.

\begin{figure}[!h]
  \centering
  \includegraphics[width=0.85\textwidth]{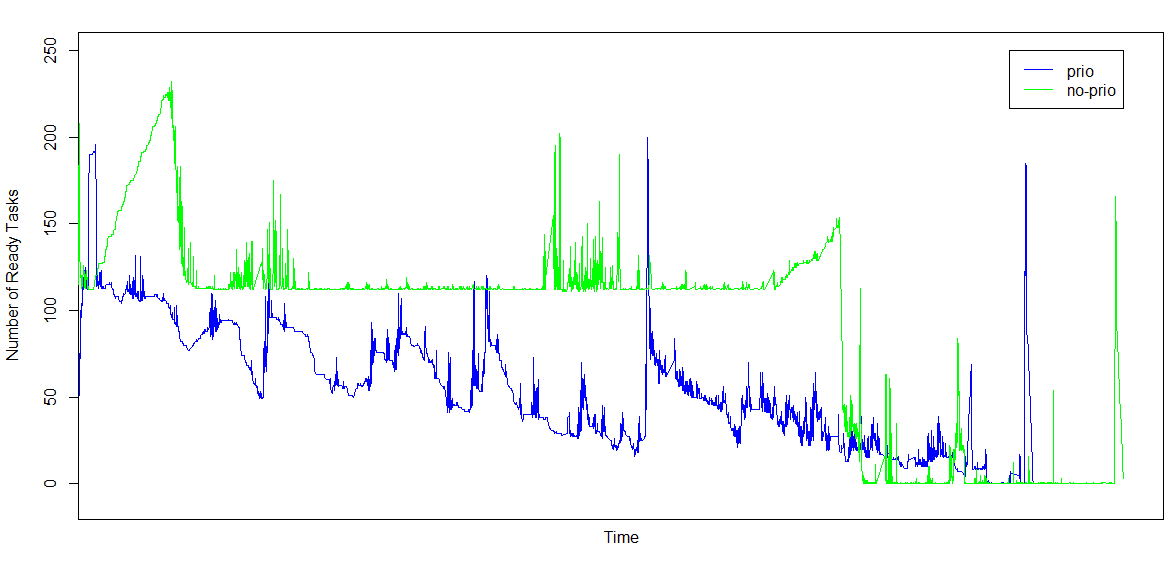}
  \caption{Evolution of the number of ready tasks (128 CEs).}
  \label{nrdy}
\end{figure}

The evolution of ready tasks in the solver is shown at Figure \ref{nrdy}. For the two strategies, we represent the number of ready tasks over time. Ready tasks are the ones already available for the workers: their dependencies have been fulfilled and they can be executed. The spike in the end comes from line 15 of Algorithm \ref{task_insertion}: for each cell component of a CE, a last task is inserted, just after a ``Wait for all'' (line 13). Moreover, line 14 corresponds to some operations that are not performed using tasks for legacy reasons.

Traversing the graph without priority unlocks more tasks at start, but in the end, not enough tasks are available to feed the workers. Our
strategy manages to keep more ready tasks available until the end.
The idle step we observe in the TS/GER part of the computation is due to the fact that with our task granularity, we execute tasks quicker
than we insert them.
Finally, we can notice that the gain achieved by using the priority
strategy is almost 10\% (14,18s versus 15,56s).

When using now 256 CEs, we notice that the sleeping time of both the solver and the TS/GER parts increases. This can be explained by the time needed to insert tasks. For example, it takes 13.6 s to insert all the tasks of the solver step (from lines 5 to line 12) and the total computing CPU time is 195.8 s (about 12.2 seconds per worker).
So, increasing the number of CEs allows more concurrency and the idle time is reduced. But with 256 CEs, the task insertion becomes so costly that it affects the global computation performance.

In conclusion of this first study in shared memory, we can say that
the way we prioritize the tasks is efficient: we manage to feed the
workers as long as possible whereas starvation was important without
priority. The task-based description truly allows to take advantage of
the irregularity of the computation. 

\subsubsection{Study with parallel workers}

In order to exploit all the CPUs without creating too much tasks, another way is the use of parallel workers. Instead of being executed on only one CPU, a task is then executed on several ones. We rely on OpenMP DO loops for the parallelization of our computational kernels, as we did in the initial OpenMP version of the code. 

\begin{figure}[!h]
  \centering
  \includegraphics[width=0.89\textwidth]{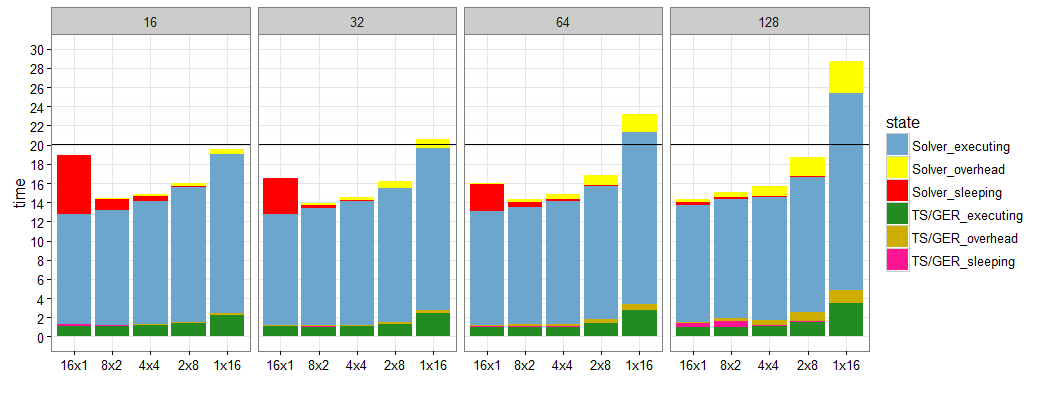}
  \caption{Time performance with different configurations for parallel workers.}
  \label{sm-par}
\end{figure}

We consider different numbers and sizes of workers (the size is the
number of cores used), from 16 workers of size 1, to one worker that
uses all the cores: so we have the relation
``({\em number\_of\_workers})$\times$({\em size\_of\_workers})=16''.
CPUs that belong to the same worker are on the same socket, except when the 2
sockets are used by one worker. We test configurations from 16 to
128 CEs. The previous ``prio'' strategy for scheduling tasks is always used in
this study.

At Figure \ref{sm-par}, we first notice that the total time spent in
computation tends to increase slightly while we use fewer and larger
workers. Indeed, the scalability of our computational kernels is not
good enough. However, this is not critical until a worker size equal
to 8. When we use the whole node and its two sockets to form only one
worker, the situation is more critical.
The overhead evolves linearly with the size of workers, while the idle
time is reduced when using parallel workers.
The best overall configuration is obtained for 32 CEs and 8 workers of size 2.

\subsubsection{Different cell repartitions in temporal levels}

At Figure \ref{ta_iter_evolution}, we highlighted the fact that the
cell repartion evolves and so the associated computation cost during
the simulation.
In Tables \ref{tl_repartition_step_1_tl_4} and
\ref{dag_info_step_2_tl_4}, we detailed two specific iterations, one
at the start of the computation (iteration 1) and the other at the
2300th iteration.
The main difference between the two cases is the computation cost
produced by the cells of low temporal levels.
Notably, for $\tau_1$ cells, the computation cost is less than 0.5\%
in the first iteration and 12.45\% in the 2300th.

In order to see the impact of this change, we compare now the
behaviour of the implementation at these two iterations. We use 3
different parallel configurations ($16 \times 1$, $8 \times 2$, $4
\times 4$) and 4 different numbers of CEs (16, 32, 64, 128). The ``prio''
strategy is always used.

\begin{figure}[!h]
  \centering
  \includegraphics[width=0.88\textwidth]{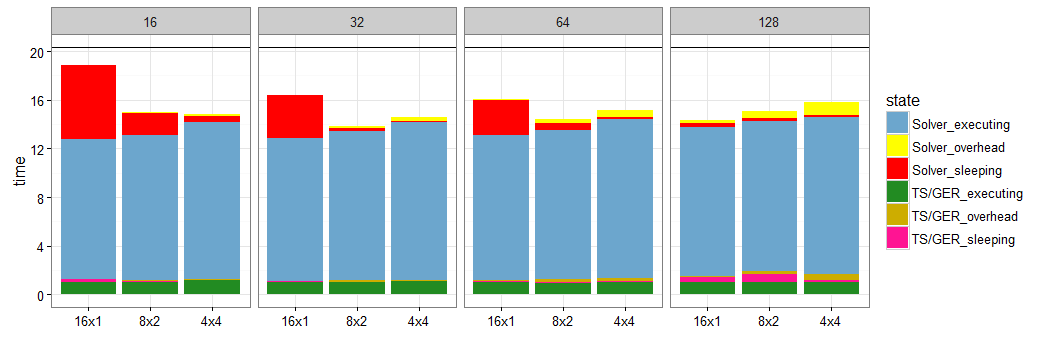}
  \caption{Time performance at first iteration ($\theta=4$).}  
  \label{step1_tl_4_perf}
\end{figure}

\begin{figure}[!h]
  \centering
  \includegraphics[width=0.88\textwidth]{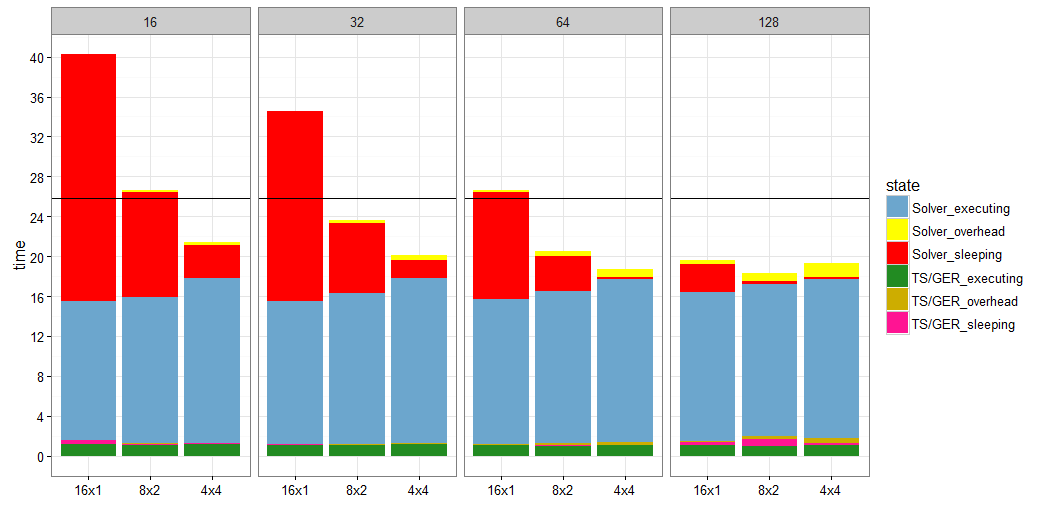}
  \caption{Time performance at 2300th iteration ($\theta=4$).}  
  \label{step2_tl_4_perf}
\end{figure}

For the first iteration, Table \ref{dag_info_step_1_tl_4} gives some
informations about the DAG that have been generated.
The \textit{Tsub} column indicates the time needed to insert all the
tasks, the \textit{elementary tasks} column indicates how many tasks
would have been generated without the usage of pack, while the
\textit{inserted packs} column indicates how many tasks were actually
inserted inside the runtime.
Figure \ref{step1_tl_4_perf} shows the performance obtained. The black
line shows the performance for the OpenMP version.
Table \ref{dag_info_step_2_tl_4} and Figure \ref{step2_tl_4_perf} give
the same informations for the 2300th iteration.

\begin{table}[h]
\centering
\begin{tabular}{| c | c | c | c |}
\hline
\#NCEs & Tsub & \# elementary tasks & \# inserted packs\\
\hline
16 & 0.7 $\pm$ 0.1 s & 23506 & 2826 \\
\hline
32 & 1.8 $\pm$ 0.1 s & 43068 & 5461 \\
\hline
64 & 3.4 $\pm$ 0.3 s & 64955 & 8696 \\
\hline
128 & 7.7 $\pm$ 0.5 s & 118633 & 16259 \\
\hline
\end{tabular}
  \caption{DAG creation informations (at first iteration).}
  \label{dag_info_step_1_tl_4}
\end{table}

\begin{table}[!h]
\centering
\begin{tabular}{| c | c | c | c |}
\hline
\#NCEs & Tsub & \# elementary tasks & \# inserted packs\\
\hline
16 & 0.6 $\pm$ 0.1 s & 34140 & 4238 \\
\hline
32 & 2.0 $\pm$ 0.1 s & 58061 & 7361 \\
\hline
64 & 3.4 $\pm$ 0.3 s & 90383 & 12006 \\
\hline
128 & 8.0 $\pm$ 0.5 s & 157605 & 21174 \\
\hline
\end{tabular}
  \caption{DAG creation informations (at 2300th iteration).}
  \label{dag_info_step_2_tl_4}
\end{table}

If we look at both Tables \ref{dag_info_step_1_tl_4} and
\ref{dag_info_step_2_tl_4}, we can notice first that the packing
strategy tends to reduce the number of tasks that the runtime has to
manage.
Increasing the number of CEs raises the number of tasks and the time
needed to submit these tasks. 
If we compare the number of tasks inserted for the two iterations, we
see that the number of tasks inserted is more important at the 2300th
iteration.

Considering now Figures \ref{step1_tl_4_perf} and
\ref{step2_tl_4_perf}, it is important to notice that the 2300th
iteration is more expensive than the first one. This is true for both
OpenMP (25.88s versus 20.32s ) and all task-based versions and this is
consistent with Figure \ref{cost_evolution_tots}.

Again, we can notice that adding more CEs reduces the synchronization
time.
Using larger workers reduces the synchronization time but has a
negative impact for the time spent executing tasks and increases the
overhead.
But if we look at the synchronization time for the 2300th iteration,
it is more important than for at the first one. With 16 CEs, only the
$4 \times 4$ configuration performs better than the OpenMP version.
Without parallel workers (configuration $16 \times 1$), the
synchronization time in this case is more important than the time
spent executing and so the task-based version is beaten by the OpenMP
one. With 32 and 64 CEs, the synchronization time is still important,
and the task-based version without parallel workers is beaten by the
OpenMP one.
This result emphasizes the gain due to the use of parallel workers :
the $4 \times 4$ configuration performs better than all the others
from 16 to 64 CEs and is only beaten by the $8 \times 2$ one with 128 CEs.
As shown by Tables \ref{dag_info_step_1_tl_4} and
\ref{dag_info_step_2_tl_4}, increasing the number of CEs implies
a more expensive task submission cost. The impact of parallel workers
on this submission cost is neglectable and their use allows to greatly
reduce the time spent in synchronization.

\subsection{A distributed memory version}

The main goal of the task-based version is to reduce the synchronization time
of the initial MPI+OpenMP version that has been higlighted by Figure \ref{tl4-n8}.
We have shown that this task-based version truly allowed to compute in a more
relaxed order.
This is notably shown in Figure \ref{gantt1} where we can see that
tasks corresponding to the first subiteration can be executed at the
end of the iteration.

To evaluate the task-based version in distributed memory, we start
with a study using the same test case than previously in shared memory. 
Then, we will use another mesh with a simpler geometry but with a greater
number of cells.
In each case, we will compare the task-based version with the initial
MPI+OpenMP version.

\subsubsection{First test case}

We still use the same test case as in the shared memory study.
We consider here experiments using from 1 to 8 MPI processes (or nodes),
with 8 to 32 CEs per process, and we configure our workers according to
3 configurations: 8$\times$2, 4$\times$4 and 2$\times$8. 
As we have seen in the shared memory experiments, using less and larger 
workers slightly tends to increase the computation time while reducing
the sleeping time. 
An increase of the number of CEs reduces the sleeping time but leads to
an augmentation of the number of tasks, and the time needed to submit all
the tasks increases too. This will be a problem for our distributed experiment 
since we have more computing capabilities and thus we expect lower time to
compute one iteration. So there is a trade-off to find between the number of
processes, the number of CEs and the number/size of workers.

\begin{figure*}[!ht]
  \includegraphics[width=0.99\textwidth]{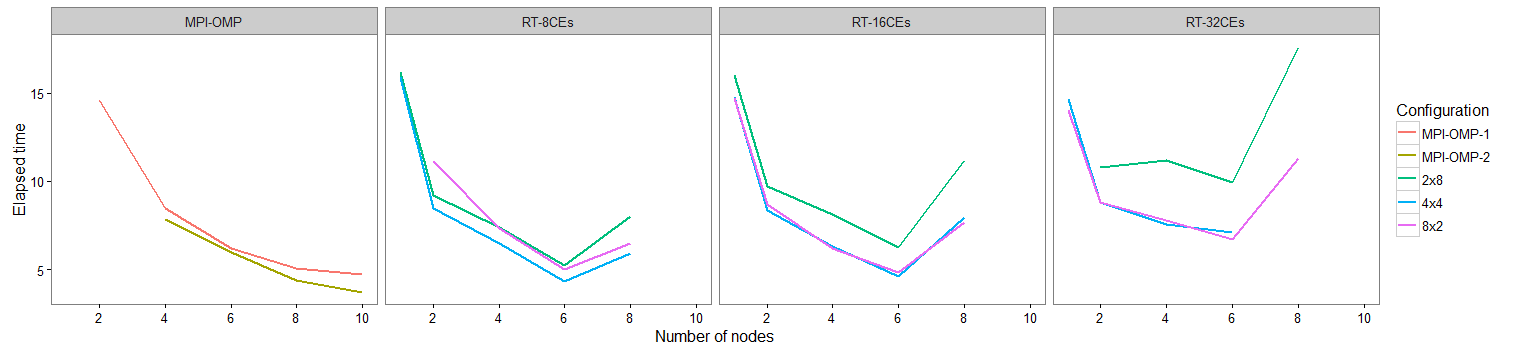}
  \caption{Elapsed time spent in the solver with different configurations (X-axis gives the number of nodes).}  
  \label{mpiperf}
\end{figure*}

\begin{figure*}[!ht]
  \includegraphics[width=0.99\textwidth]{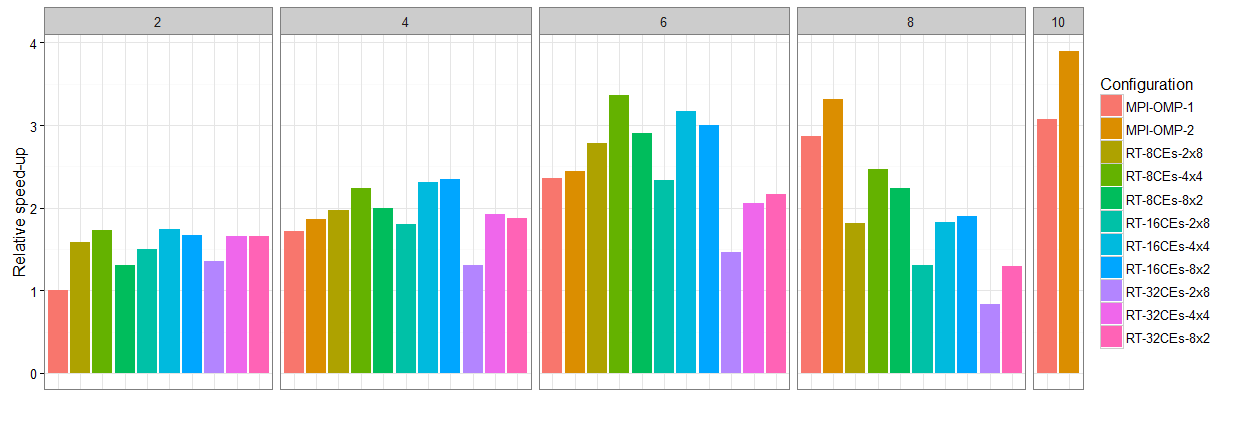}
  \caption{For a number of nodes varying from 2 to 8, relative speed-up compared to the MPI-OMP-1 version with two nodes.}  
  \label{relspeed_up}
\end{figure*}

Each process is in charge of a portion of the mesh and the domain decomposition
takes into account the cost of cells.
In the Figures \ref{mpiperf} and \ref{relspeed_up}, we denote by \textit{MPI-OMP-<ppn>}
the computations made with the MPI+OpenMP version with \textit{<ppn>} processes per node.
Computations made with the task-based version are denoted \textit{RT-<n>CEs-<configuration>}
with \textit{<n>} being the number of CEs per node and \textit{<configuration>} being the way
the workers are configured (\textit{number of workers $\times$ worker size}).

Figure \ref{mpiperf} shows the different elapsed times for different configurations tested:
the leftmost diagram corresponds to the MPI+OpenMP versions with a number of nodes varying from 2 to 10,
and the three others correspond to the task-based versions with a varying number of CEs per node
and with a number of nodes varying from 1 to 8.

\begin{table*}[!ht]
\centering
\begin{tabular}{| l | c | c | c | c | c | c |}
\hline
& TS/GER idle & TS/GER executing & Solver idle & Solver executing & Submission time & Elapsed time \\
\hline
Process \#1 & 0.338 & 0.197 & 1.440 & 2.356 & 0.608 & 4.331  \\
\hline
Process \#2 & 0.357 & 0.174 & 1.559 & 2.241 & 1.301  & 4.331  \\
\hline
Process \#3 & 0.351 & 0.184 & 1.401 & 2.395 & 0.937 & 4.331  \\
\hline
Process \#4 & 0.293 & 0.243 & 1.790 & 2.005 & 0.218 & 4.331  \\
\hline
Process \#5 & 0.310 & 0.216 & 1.599 & 2.205 & 0.599 & 4.331  \\
\hline
Process \#6 & 0.344 & 0.191 & 1.627 & 2.169 & 1.259 & 4.331  \\
\hline
\end{tabular}
  \caption{Average time (in seconds) for each process. We use 6 processes, 8 CEs per process, 4 workers and 4 cores per worker.}
  \label{nsla6}
\end{table*}

If we consider first the task-based version, we can see that the best performance (4,331s) is obtained with 6 processes,
8 CEs per process, each one configured as 4$\times$4 (4 workers and 4 cores per worker).
This case is detailed in Table \ref{nsla6}.
The load balancing is good with a difference of 13\% for executing time (including \textit{TS/GER} and \textit{Solver} parts)
between Process 1 and Process 4.
However, the idle time is important: with only 2 CEs per worker, it is not easy to exploit enough asynchronism and
communication-computation overlapping.

The first diagram of Figure \ref{mpiperf} shows the elapsed time for the initial MPI+OpenMP version of the aerodynamic
solver. We consider here 2 configurations, one with one process per node, and one with one process per socket.
Unfortunately, due to memory limitations, it has been impossible to run the test case with a larger number of processes
per node.
We can see that the absolute best performance is obtained for this version with 2 processes per node and 10 nodes. 
But when we consider a number of nodes for which the task-based version is competitive (at most 6 nodes), the task-based
version gives a better result than the MPI+OpenMP one (4.33s versus 5.98s for 2 processes per node and 6 nodes).

However, with 8 MPI processes, we don't reduce the elapsed time in the solver for the task-based version.
This is due to the number of CEs which varies from 64 for 8 CEs per process to 256 for 32 CEs per process.
When we use more nodes, the computation time by node decreases, but the overhead stays almost the same.
This is the main reason why we cannot use an important number of CEs and load balancing is also a concern.

Another illustration of these results is given at Figure \ref{relspeed_up}. It shows now the relative speed-up of the
task-based versions compared to the MPI+OpenMP version using 2 nodes and 1 process per node.
Each diagram corresponds to a given number of nodes (from 2 to 8) and each configuration is represented by a specific color.
This way, it is easy to compare performances for a given number of nodes and the scalability for a given configuration.

To conclude this first study in distributed memory, we can say that, in fact, the size of the considered test case is a little bit
tiny to overcome the overhead induced by the task-based parallelization. Moreover, it's clear that the way the task-based version
is configured has a strong impact on performance.

\subsubsection{A larger test-case}

In this section, we evaluate the behavior of the task-based version
on a larger problem and with a larger number of nodes; we compare
its performances to the ones of the initial MPI+OpenMP version. 

The test case is still a take-off blast wave propagation with a simpler geometry but with 80 M cells.
Due to the simpler mesh geometry, only 4 temporal levels are present at most in this computation ($\theta=3$).
Since this test case is larger, we expect to be able to run with more processes: the overhead proportion should
be lesser than for the previous test case and should allow the task-based version to be more competitive.

For this experiment, we used a cluster with 20 cores per node (2 Ivybridge processors with 10 cores each).

We run the initial MPI+OpenMP version with one process (named MPI-OMP-1) or two processes (named MPI-OMP-2)
per node, and we fix a $4 \times 5$ configuration for the task-based version (4 workers each using 5 cores).
This task-based version is performed by using 10 CEs per process (named RT-10CEs) or 20 CEs per process (named RT-20CEs).

In order to be able to run the computation without memory swapping, we must use at least 16 nodes\footnote{The cluster used
did not allow to swap.}.
Concerning the MPI+OpenMP version using 16 nodes, only the computation with one process per node could be achieved due to a
memory consumption problem when using 2 processes per node.

\begin{figure}[!h]
  \centering
  \includegraphics[width=0.63\textwidth]{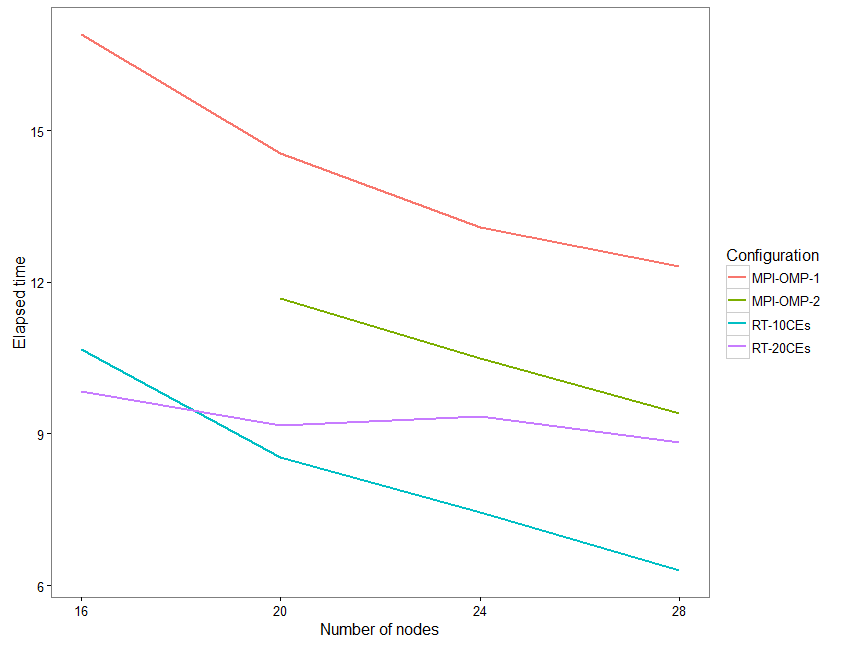}
  \caption{Elapsed time spent in the solver with different configurations for the 80M cells test case.}
  \label{lc_time}
\end{figure}

\begin{figure}[!h]
  \centering
  \includegraphics[width=0.99\textwidth]{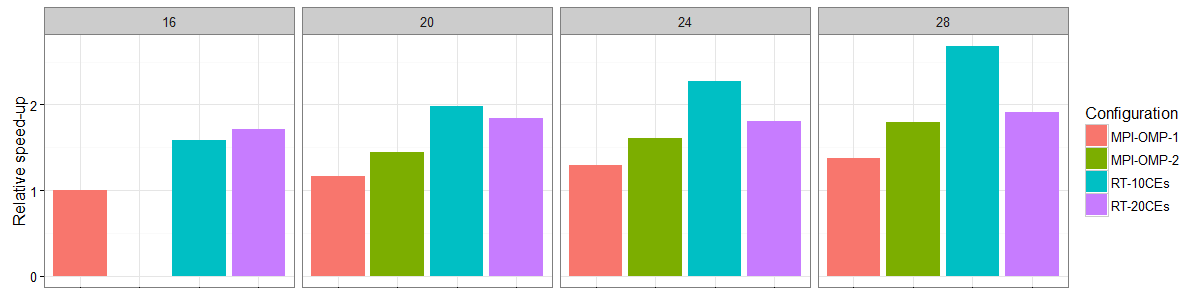}
  \caption{For a number of nodes varying from 16 to 28, relative speed-up compared to the MPI-OMP-1 version with 16 nodes.}
  \label{lc_speed-up}
\end{figure}

The experiment is done for 16, 20, 24 and 28 nodes.
Figure \ref{lc_time} shows the elapsed time for each version, and Figure \ref{lc_speed-up} shows the relative speed-up compared
to MPI-OMP-1 version using 16 nodes.

Regarding the MPI+OpenMP version, the configuration using 2 processes per node is clearly better than the one with only one
process per node.

For this larger test case, the task-based version using StarPU is more efficient than the MPI+OpenMP version.
With 16 nodes, the best task-based configuration (with 20 CEs per node) is 70\% faster than the best MPI+OpenMP version
(with 1 process per node).
With 28 nodes, this gain is still 49\%, but here the best task-based configuration uses 10 CEs while the best MPI+OpenMP
version uses 2 processes per node.
We notice that the task-based version using 10 CEs per node scales better than when using 20 CEs per node. Indeed, the
computation granularity becomes more and more problematic as the number of nodes increases (the number of cells per CE
is lower when using 20 CEs per node).

\begin{figure}[!h]
  \centering
  \includegraphics[width=0.99\textwidth]{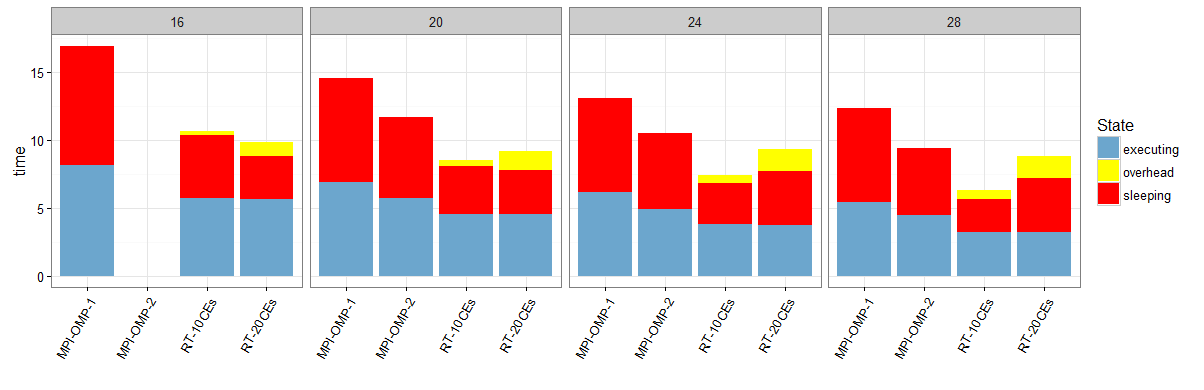}
  \caption{Time spent in each state for the 80M cells test case with different configurations.}
  \label{lc_proportion}
\end{figure}

Figure \ref{lc_proportion} gives some details about the time spent in the different states for the different configurations
and for the different numbers of nodes.

The Y-axis shows the total time for an iteration while the bars are
colored according to the proportion spent in each state: in red for
the sleeping time, in blue for the computing time and in yellow for
the overhead time.

For the MPI+OpenMP version, we notice that using 2 processes per node
reduces significantly the time spent in computing. The proportion of
time spent in the sleeping state remains around 50\% in all the cases.

If we compare the two configurations used in the task-based version,
the best one for 16 nodes is the one using 20 CEs per node. For this
number of nodes, increasing the number of CEs allows to reduce the
time spent in the sleeping state. However, the overhead is more important
than with 10 CEs per node.
If now we use more nodes, the best version becomes the one using 10 CEs per
node. The overhead implied by the increasing number of CEs become more
important. Furthermore, by adding more nodes, we increase the total
number of CEs and each CE contains less and less cells.\\

It would be nice to be able to limit this overhead in order to obtain
a better gain when using the task-based version. The easier way to get
more asynchronism is to increase the number of CEs, but this also
increases the associated overhead which currently limits the potential
gain.
As we can see, there is a lot of parameters (number of CEs, strategy for
task priorities, configuration of workers, number of processes) that can
be adjusted for this tasked-based version and they may have a strong impact
on performance. Finding the best values of these parameters for a given
parallel platform is clearly a critical issue.
Moreover, this configuration should be able to evolve during the computation 
because of cell distribution changes in the temporal level classes.
So, having a way to change the worker configuration and the number of 
CEs on-the-fly and automatically could be of great interest in order to 
achieve the best possible performances.

\section{Conclusion and perspectives}
\label{conclusion}

In this paper, we have described a task-based distributed version of an aerodynamic solver with adaptive time step.
We have validated our implementation with several real-life industrial cases. 
When under the right conditions (i.e. when the overhead induced by the runtime system is not too high), the implementation
shows very promising results. The parallelism offered by the task-based paradigm allows a better exploitation of the computational
resources.
On the largest test case, the task-based version performs much better than the initial MPI+OpenMP version.
However, it is not possible to remove all synchronizations: if we use too much CEs per process, the overhead induced is too high.
Some code analysis should be performed in order to see if it is possible to reduce this overhead.\\

It is clear that exploiting parallel tasks is useful by reducing the number of workers, but this point can be limited by the scalability
of the computational kernels.
Concerning the way we exploit parallel tasks, we tested only the simple case of multiple workers of the same size. It is possible to explore
more heterogeneous configurations (e.g. 1$\times$8+2*2$\times$4: one worker of size 8, and two of size 4). It could even be possible to modify
the configuration on the fly regarding the current status of computations. To overcome the starvation seen at the end of the iteration, one
possible strategy is to gather all CPUs into one worker when there is few ready tasks.
Context resizing is now possible in StarPU and first experimentations have been described in \cite{hugo:hal-00824514}.
If the effort is made to get GPU versions of the kernels, 
more heterogeneity could be exploited. 
After the refactorization of the code, kernels now follow specific patterns.
So, it may be possible to exploit this fact by developing
a source-to-source compiler that would translate a CPU version
of a kernel into a GPU one.

Right now, the decomposition into CEs is fixed at the start of the computation, but the temporal level of cells evolves during the computation
between iterations. Having the possibility to reshape the CEs could lead to an improvement: it should be possible to generate a more optimized
DAG with CEs that take into account the temporal levels of cells.
Another interest for the CE reshape could be the improvement of load balancing for the distributed version. This could be done by exchanging CEs
between processes after an iteration has been completed and before the next one.
Furthermore, if we manage to remove the hard synchronization we have at the end of general iterations, we can expect to pipeline iterations while applying an asynchronous load balancing scheme.

In order to optimize the parameters of the configurations in the task-based version without making systematic tests, it could be interesting
to port the code into StarPU-SimGrid~\cite{stanisic}. SimGrid is a versatile simulator for distributed systems that allows to obtain accurate
performance predictions. By collecting data during actual computations, it is possible to model the way kernels behave under different circumstances
and then inject those data in the simulator. Some features would need to be added in StarPU-SimGrid (e.g. support for parallel tasks), but it could
be a tool of great interest to tune efficiently the scheduling priorities or the worker configuration parameters.

From an application point of view, we intend now to develop a task-based parallelization of the intersection mechanism of FLUSEPA in order to exploit more asynchronism in the whole application when considering booster separation simulations.
With the MPI+OpenMP version, we are already able to exploit asynchronism by computing mesh intersections in advance. However, since the operations
are handled by different processes, the application of the intersection process is costly and involves a lot of communications. So, we intend to
co-schedule the tasks related to intersection computations and to aerodynamic computations on the same processes exploiting data locality.\\

This work is a proof of concept showing that a task-based parallelization over a runtime system can be a good approach on the route towards efficient large scale simulations on forthcoming post-petascale supercomputers.
Achieving good performances could be easier from a task-based description: adjusting the grain of parallelism will be possible and in our case
it can be done by choosing the number of CEs or the configuration used for parallel workers.
Overlapping of communication by computation can also be taken to another level. First, the fine description of dependencies allows to remove synchronizations. Also, it is possible to have scheduling strategies that favor communication/computation overlapping by selecting the right
tasks that will generate communications as soon as possible. 
However, the programming effort in order to get a task-based version is important and relies on very specific libraries which can be risky to
ensure the continued existence of an industrial code.
Thoses problems may be reduced by the rise of programming standards like OpenMP 4.5~\cite{openmp15} which allows to do things similar to STF
paradigm through the \textit{depend} clause.
Moreover, some efforts are currently done to bridge the gap between OpenMP 4.0 and the native runtimes. We can mention KSTAR~\cite{agullo:hal-01372022} tool which is a source-to-source compiler that translates OpenMP directives into calls to task-based runtime system APIs.

Another thing to consider is the development of runtime-ready frameworks for different classes of applications : since there are similarities between applications that use the same numerical methods, this fact could be exploited. This way, physicists could concentrate on coding models while the performance aspects would be handled by the framework.

\bibliographystyle{elsarticle-num}
\bibliography{pdsec}

\begin{thebibliography}{10}
\expandafter\ifx\csname url\endcsname\relax
  \def\url#1{\texttt{#1}}\fi
\expandafter\ifx\csname urlprefix\endcsname\relax\def\urlprefix{URL }\fi
\expandafter\ifx\csname href\endcsname\relax
  \def\href#1#2{#2} \def\path#1{#1}\fi

\bibitem{forum_mpi:_1994}
M.~P. Forum, {MPI: A Message-Passing Interface Standard}, Tech. rep.,
  Knoxville, TN, USA (1994).

\bibitem{dagum_openmp:_1998}
L.~Dagum, R.~Menon, {OpenMP}: an industry standard {API} for shared-memory
  programming, Computational Science \& Engineering, {IEEE} 5~(1) (1998)
  46--55.

\bibitem{AugThiNamWac11CCPE}
C.~Augonnet, S.~Thibault, R.~Namyst, P.-A. Wacrenier,
  \href{http://hal.inria.fr/inria-00550877}{{StarPU: A Unified Platform for
  Task Scheduling on Heterogeneous Multicore Architectures}}, Concurrency and
  Computation: Practice and Experience, Special Issue: Euro-Par 2009 23 (2011)
  187--198.
\newblock \href {http://dx.doi.org/10.1002/cpe.1631}
  {\path{doi:10.1002/cpe.1631}}.
\newline\urlprefix\url{http://hal.inria.fr/inria-00550877}

\bibitem{bosilca_parsec:_2013}
G.~Bosilca, A.~Bouteiller, A.~Danalis, M.~Faverge, T.~Hérault, J.~Dongarra,
  \href{http://hal.inria.fr/hal-00930217}{{PaRSEC}: A programming paradigm
  exploiting heterogeneity for enhancing scalability}, Computing in Science and
  Engineering 99 (2013) 1.
\newblock \href {http://dx.doi.org/10.1109/MCSE.2013.98}
  {\path{doi:10.1109/MCSE.2013.98}}.
\newline\urlprefix\url{http://hal.inria.fr/hal-00930217}

\bibitem{hugo:hal-00824514}
A.~Hugo, A.~Guermouche, R.~Namyst, P.-A. Wacrenier,
  \href{http://hal.inria.fr/hal-00824514}{{Composing multiple StarPU
  applications over heterogeneous machines: a supervised approach}}, in: {Third
  International Workshop on Accelerators and Hybrid Exascale Systems}, Boston,
  USA, 2013.
\newline\urlprefix\url{http://hal.inria.fr/hal-00824514}

\bibitem{brenner1991three}
P.~Brenner, Three-dimensional aerodynamics with moving bodies applied to solid
  propellant, in: 27th Joint Propulsion Conference, 1991.

\bibitem{kleb_temporal_1992}
W.~L. Kleb, J.~T. Batina, M.~H. Williams,
  \href{http://arc.aiaa.org/doi/pdf/10.2514/3.11169}{{Temporal adaptive
  Euler/Navier-Stokes algorithm involving unstructured dynamic meshes}}, {AIAA}
  journal 30~(8) (1992) 1980--1985.
\newline\urlprefix\url{http://arc.aiaa.org/doi/pdf/10.2514/3.11169}

\bibitem{lohner_finite_1985}
R.~Löhner, K.~Morgan, J.~Peraire, O.~A. Zienkiewicz, Finite element methods
  for high speed flows, University College of Swansea Institute for Numerical
  Methods in Engineering, 1985.

\bibitem{pervaiz_temporal_1988}
M.~M. Pervaiz, J.~R. Baron,
  \href{http://onlinelibrary.wiley.com/doi/10.1002/cnm.1630040114/abstract}{Temporal
  and spatial adaptive algorithm for reacting flows}, Communications in Applied
  Numerical Methods 4~(1) (1988) 97--111.
\newblock \href {http://dx.doi.org/10.1002/cnm.1630040114}
  {\path{doi:10.1002/cnm.1630040114}}.
\newline\urlprefix\url{http://onlinelibrary.wiley.com/doi/10.1002/cnm.1630040114/abstract}

\bibitem{mypdsec}
J.~M. Couteyen~Carpaye, J.~Roman, P.~Brenner, {Towards an efficient Task-based
  Parallelization over a Runtime System of an Explicit Finite-Volume CFD Code
  with Adaptive Time Stepping}, in: {International Parallel and Distributed
  Processing Symposium}, PDSEC'2016 workshop of IPDPS, Chicago, IL, United
  States, 2016.
\newblock \href {http://dx.doi.org/10.1109/IPDPSW.2016.125}
  {\path{doi:10.1109/IPDPSW.2016.125}}.

\bibitem{versteeg_introduction_2007}
H.~K. Versteeg, W.~Malalasekera, An Introduction to Computational Fluid
  Dynamics: The Finite Volume Method, Pearson Education, 2007.

\bibitem{germano1999rans}
M.~Germano, From {RANS} to {DNS}: towards a bridging model, in: Direct and
  Large-Eddy Simulation III, Springer, 1999, pp. 225--236.

\bibitem{godounov_resolution_1979}
S.~Godounov, A.~Zabrodine, M.~Ivanov, A.~Kraiko, G.~Prokopov, Résolution
  numérique des problemes multidimensionnels de la dynamique des gaz, Editions
  Mir, 1979.

\bibitem{hirt_arbitrary_1974}
C.~W. Hirt, A.~A. Amsden, J.~L. Cook,
  \href{http://www.sciencedirect.com/science/article/pii/0021999174900515}{An
  arbitrary lagrangian-eulerian computing method for all flow speeds}, Journal
  of Computational Physics 14~(3) (1974) 227--253.
\newblock \href {http://dx.doi.org/10.1016/0021-9991(74)90051-5}
  {\path{doi:10.1016/0021-9991(74)90051-5}}.
\newline\urlprefix\url{http://www.sciencedirect.com/science/article/pii/0021999174900515}

\bibitem{gillyboeuf1995two}
J.-P. Gillyboeuf, P.~Mansuy, S.~Pavsic, Two new {Chimera} methods: application
  to missile separation, Tir{\'e} {\`a} part- Office national d'{\'e}tudes et
  de recherches aerospatiales.

\bibitem{kao1994grid}
K.-H. Kao, M.-S. Liou, C.-Y. Chow, {Grid adaptation using Chimera composite
  overlapping meshes}, AIAA journal 32~(5) (1994) 942--949.

\bibitem{brenner_simulation_1998}
P.~Brenner, J.-M. Carrat, M.~Pollet, Simulation d'interactions
  a{\'e}rodynamiques instationnaires autour de plusieurs corps en mouvement
  relatif, in: {AAAF} : Les interactions en a{\'e}rodynamique, 1998.

\bibitem{ompss}
A.~{Duran}, J.~M. {Perez}, R.~M. {Ayguad\'e}, E. amd~{Badia}, J.~{Labarta},
  Extending the {OpenMP} tasking model to allow dependent tasks, in: {OpenMP}
  in a New Era of Parallelism, 4th International Workshop, {IWOMP} 2008,
  Lecture Notes in Computer Science 5004:111-122, West Lafayette, IN, 2008.

\bibitem{budimlic_concurrent_2010}
Z.~Budimlić, M.~Burke, V.~Cavé, K.~Knobe, G.~Lowney, R.~Newton, J.~Palsberg,
  D.~Peixotto, V.~Sarkar, F.~Schlimbach, S.~Taşirlar,
  \href{http://dl.acm.org/citation.cfm?id=1938482.1938486}{Concurrent
  collections}, Sci. Program. 18~(3) (2010) 203--217.
\newline\urlprefix\url{http://dl.acm.org/citation.cfm?id=1938482.1938486}

\bibitem{bauer2014legion}
M.~E. Bauer, Legion: Programming distributed heterogeneous architectures with
  logical regions, Ph.D. thesis, Stanford University (2014).

\bibitem{chan_supermatrix_2007}
E.~Chan, E.~S. Quintana-Orti, G.~Gregorio Quintana-Orti, R.~van~de Geijn,
  Supermatrix out-of-order scheduling of matrix operations for {SMP} and
  multi-core architectures, in: Nineteenth Annual {ACM} Symposium on Parallel
  Algorithms and Architectures {SPAA}'07, 2007, pp. 116--125.

\bibitem{bosilca_flexible_2011}
G.~Bosilca, A.~Bouteiller, A.~Danalis, M.~Faverge, A.~Haidar, T.~Herault,
  J.~Kurzak, J.~Langou, P.~Lemarinier, H.~Ltaief, P.~Luszczek, A.~{YarKhan},
  J.~Dongarra, Flexible development of dense linear algebra algorithms on
  massively parallel architectures with {DPLASMA}, in: 12th {IEEE}
  International Workshop on Parallel and Distributed Scientific and Engineering
  Computing ({PDSEC}'11), 2011.

\bibitem{lacoste_taking_2014}
X.~Lacoste, M.~Faverge, P.~Ramet, S.~Thibault, G.~Bosilca,
  \href{http://hal.inria.fr/hal-00925017}{Taking advantage of hybrid systems
  for sparse direct solvers via task-based runtimes}, in: Proceedings of the
  {IEEE} International Symposium on Parallel \& Distributed Processing
  Workshops and Phd Forum ({IPDPSW}'14), {HCW} 2014, 2014.
\newline\urlprefix\url{http://hal.inria.fr/hal-00925017}

\bibitem{agullo2014task}
E.~Agullo, B.~Bramas, O.~Coulaud, E.~Darve, M.~Messner, T.~Takahashi,
  Task-based {FMM} for multicore architectures, SIAM Journal on Scientific
  Computing 36~(1) (2014) C66--C93.

\bibitem{cosnard1995automatic}
M.~Cosnard, M.~Loi, Automatic task graph generation techniques, in: System
  Sciences, 1995. Proceedings of the Twenty-Eighth Hawaii International
  Conference on, Vol.~2, IEEE, 1995, pp. 113--122.

\bibitem{allen2002optimizing}
R.~Allen, K.~Kennedy, Optimizing compilers for modern architectures: a
  dependence-based approach, Vol. 289, Morgan Kaufmann San Francisco, 2002.

\bibitem{cojean:hal-01181135}
T.~Cojean, A.~Guermouche, A.~Hugo, R.~Namyst, P.-A. Wacrenier, {Exploiting
  two-level parallelism by aggregating computing resources in task-based
  applications over accelerator-based machines},
  https://hal.inria.fr/hal-01181135 (2015).

\bibitem{pellegrini1996scotch}
F.~Pellegrini, J.~Roman, Scotch: A software package for static mapping by dual
  recursive bipartitioning of process and architecture graphs, in:
  High-Performance Computing and Networking, Springer, 1996, pp. 493--498.

\bibitem{stanisic}
L.~Stanisic, S.~Thibault, A.~Legrand, B.~Videau, J.-F. M{\'e}haut,
  \href{https://hal.inria.fr/hal-01147997}{{Faithful Performance Prediction of
  a Dynamic Task-Based Runtime System for Heterogeneous Multi-Core
  Architectures}}, {Concurrency and Computation: Practice and Experience}
  (2015) 16\href {http://dx.doi.org/10.1002/cpe} {\path{doi:10.1002/cpe}}.
\newline\urlprefix\url{https://hal.inria.fr/hal-01147997}

\bibitem{openmp15}
{OpenMP Architecture Review Board},
  \href{http://www.openmp.org/mp-documents/openmp-4.5.pdf}{{OpenMP} application
  program interface version 4.5} (2015).
\newline\urlprefix\url{http://www.openmp.org/mp-documents/openmp-4.5.pdf}

\bibitem{agullo:hal-01372022}
E.~Agullo, O.~Aumage, B.~Bramas, O.~Coulaud, S.~Pitoiset,
  \href{https://hal.inria.fr/hal-01372022}{{Bridging the gap between OpenMP 4.0
  and native runtime systems for the fast multipole method}}, Research Report
  RR-8953, {Inria} (Mar. 2016).
\newline\urlprefix\url{https://hal.inria.fr/hal-01372022}

\end{thebibliography}

\end{document}